\documentclass[fleqn,usenatbib]{mnras}

\usepackage{newtxtext,newtxmath}
\usepackage{subcaption}
\usepackage{rotating}
\usepackage[T1]{fontenc}

\DeclareRobustCommand{\VAN}[3]{#2}
\let\VANthebibliography\thebibliography
\def\thebibliography{\DeclareRobustCommand{\VAN}[3]{##3}\VANthebibliography}

\usepackage{graphicx}	% Including figure files
\usepackage{amsmath}
\usepackage{gensymb}
\usepackage{siunitx}% Advanced maths commands
\usepackage{subcaption}

\newcommand{\kepler}{{\it Kepler}}

\newcommand{\tess}{{\it TESS}}
\newcommand{\PLATO}{{\it PLATO}}
\newcommand{\plato}{{\it PLATO}}

\newcommand{\mjup}{\mbox{M\textsubscript{J}}}

\newcommand{\rsun}{\mbox{R$_{\odot}$}}

\newcommand{\rearth}{R$_{\oplus}$}
\newcommand{\mearth}{M$_{\oplus}$}

\title[Viewing the PLATO Field Through TESS]{Viewing the PLATO LOPS2 Field Through the Lenses of TESS}

\author[Y. N. E. Eschen et al.]{Yoshi Nike Emilia Eschen,$^{1}$
Daniel Bayliss,$^{1}$
Thomas G. Wilson,$^{1}$
Michelle Kunimoto,$^{2}$
Ingrid Pelisoli,$^{1}$
\newauthor
Toby Rodel$^{3}$
\\
$^{1}$Department of Physics, University of Warwick, Gibbet Hill Road, Coventry, CV4 7AL, UK,\\
$^{2}$Department of Physics and Astronomy, University of British Columbia, 6224 Agricultural Road, Vancouver, BC V6T 1Z1, Canada,\\
$^{3}$Astrophysics Research Centre, School of Mathematics and Physics, Queen’s University Belfast, Belfast, BT7 1NN, UK}

\date{Accepted XXX. Received YYY; in original form ZZZ}

\pubyear{2024}

\begin{document}
\label{firstpage}
\pagerange{\pageref{firstpage}--\pageref{lastpage}}
\maketitle

\begin{abstract}
\plato\ will begin observing stars in its Southern Field (LOPS2) after its launch in late 2026.  By this time, \tess\ will have observed the stars in LOPS2 for at least four years. We find that by 2025, on average each star in the \plato\ field will have been monitored for 330 days by \tess, with a subset of stars in the \tess\ continuous viewing zone having over 1000 days of monitoring.  There are currently 101 known transiting exoplanets in the LOPS2 field, with 36 of these residing in multiplanet systems.  The LOPS2 field also contains more than 500 \tess\ planet candidate systems, 64 exoplanets discovered by radial velocity only, over 1000 bright (V$<$13) eclipsing binary systems, 7 transiting brown dwarf systems, and 2 bright white dwarfs (G$<$13).  We calculate  \tess\ and \plato\ sensitivities to detecting transits for the bright FGK stars that make up the \plato\ LOPS2 P1 sample.  We find that \tess\ should have discovered almost all transiting giant planets out to approximately 30\,d within the LOPS2 field, and out to approximately 100\,d for the regions of the LOPS2 field within the \tess\ CVZ ($\sim20$ per cent of the LOPS2 field).  However, we find that for smaller radius planets in the range 1\,--\,4\,\rearth\  \plato\ will have significantly better sensitivity, and these are likely to make up the bulk of new \plato\ discoveries.   
\end{abstract}

%\begin{keywords}
%keyword1 -- keyword2 -- keyword3
%\end{keywords}

%%%%%%%%%%%%%%%%%%%%%%%%%%%%%%%%%%%%%%%%%%%%%%%%%%

%%%%%%%%%%%%%%%%% BODY OF PAPER %%%%%%%%%%%%%%%%%%
\section{Introduction}
\label{sec:intro}
Following the discovery of the first exoplanets in the 1990s \citep{1992Natur.355..145W, 1995Natur.378..355M}, over 5000 exoplanets have been discovered using various techniques \citep[][accessed on 9 October 2024]{PSCompPars}.  The transit method \citep{sackett1999, seager, 2010exop.book...55W} is, to-date, the most successful technique for discovering exoplanets, currently accounting for 4274 of the known exoplanets \citep[][accessed on 31 July 2024]{PSCompPars}.  A large fraction of this success has been due to wide-field space-based photometric surveys such as $CoRoT$ \citep{corot}, $Kepler$ \citep{kepler}, $K2$ \citep{k2}, and \tess\ \citep{2015JATIS...1a4003R}. \\
A new European Space Agency (ESA) mission searching for transiting exoplanets is scheduled for launch in late 2026: Planetary Transits and Oscillations of Stars \citep[\plato;][]{plato_2024}.  One of the main goals of \plato\ is to discover terrestrial planets in the habitable zones of solar-like stars.  To do this \plato\ will observe stars with multiple cameras (between 6 and 24 depending on the location of the star in the field), and is estimated to achieve a precision of $\sim$50 ppm in one hour for a star at V=11\,mag \citep{boerner2024plato_noise}.  The initial field to be observed by \plato\ has now been confirmed, and is a field in the Southern Ecliptic Hemisphere known as the ``LOPS2'' field \citep{platofield}.  \plato\ will observe the LOPS2 field for at least the first two years of the mission \citep{plato_2024}.\\
The Transiting Exoplanet Survey Satellite \citep[\tess;][]{2015JATIS...1a4003R} has been conducting an all-sky photometric survey for transiting exoplanets since it was launched in 2018.  Since \tess\ has now observed the Southern Ecliptic Hemisphere three times, and will continue to observe it in the future, all of the stars in the \plato\ LOPS2 field will have a significant amount of photometric data from the \tess\ mission prior to the launch of \plato.  In this work, we investigate what this \tess\ data can tell us about the stars in the LOPS2 field, and what impact that will have for discovering exoplanets with the \plato\ mission.\\
In this paper we outline the key aspects of the \tess\ and \plato\ missions in \autoref{sec:missions}.  We then describe our methodology for determining the sensitivities of \tess\ and \plato\ to discovering transiting exoplanets in the LOPS2 field in \autoref{sec:method}.  
% cover our methods in \autoref{sec:methods} including CDPP calculations and the creation of sensitivity maps. 
In \autoref{sec:results} we present and discuss our results including transiting planets and planet candidates, planets only detected by radial velocity as well as eclipsing binaries and bright white dwarf systems in the \plato\, LOPS2 field. This section also includes our results from the precision calculations and sensitivity maps. We summarise our conclusions in \autoref{sec:conclusions}.\\
\section{The TESS and PLATO Missions}
\label{sec:missions}

\subsection{TESS}
\label{sec:tess}
\tess\ \citep{2015JATIS...1a4003R} is a NASA Astrophysics Explorer mission led and operated by MIT in Cambridge, Massachusetts, and managed by NASA’s Goddard Space Flight Center.  \tess\, is in a 13.7\,d elliptical orbit in a 2:1 resonance with the moon.  It is performing an all-sky survey by observing sectors of \SI{2300}{\deg\squared} for 27.4\,d each (two orbits).  These sectors typically tile the ecliptic hemispheres, with overlapping regions that result in areas of longer duration coverage (see \autoref{fig:plato_tess_field}).  Around the ecliptic poles this includes the continuous viewing zones (CVZ) which are monitored for one year in each hemisphere.  Temporal data gaps occur between \tess\, orbits and sectors, and during any periods of technical problems with the cameras or spacecraft.  For a typical star in the CVZ this results in approximately 20 per cent less duration than a truly continuous coverage \citep{TIARA}.\\
\tess\ consists of four f/1.4 lenses (\SI{10}{\centi\meter} effective aperture), each coupled to an array of four 2K$\times$4K pixel CCDs (with 2K$\times$2K of imaging pixels) with a pixel scale of 21 arcsec/pixel. Data is read out every 2 seconds and summed to 20 seconds or 2 minutes for postage stamps which are images (11$\times$11 pixels) of selected stars containing the star and the pixels around it,  to 200 seconds (in Extended Mission 2), 10 minutes (in Extended Mission 1) or 30 minutes (in Primary Mission) for Full-Frame Images. Data processing is used to mitigate the effects of cosmic rays and stray light from Earth or the Moon \citep{tess_handbook}. Information on cosmic rays and stray light are made publicly available for each sector. \tess\ data is then further processed on the ground by two main pipelines: the Science Processing and Operations Centre \citep[SPOC;][]{spoc, spoc_caldwell} and the Quick Look Pipeline \citep[QLP;][]{qlp, qlp_2, qlp3, qlp4}. SPOC processes the sample of pre-selected stars at the 2-minute cadence \citep{2015JATIS...1a4003R} as well as up to 160,000 FFI lightcurves per sector using a selection function outlined in \citet{spoc_caldwell} using simple aperture photometry.  QLP processes around 1,000,000 light curves per sector including all stars down to T=13.5\,mag  and M dwarfs as faint as T=15\,mag using multi-aperture photometry.\\
\tess\ has been extremely successful in discovering transiting exoplanets, beginning with pi Mensae\,c, a Neptune-sized exoplanet transiting a very bright (V=5.7) star in a 6.3\,d orbit \citep{pimen2,PiMen}.  So far \tess\ has discovered 543 confirmed planets \citep[][accessed on 1 August 2024]{PSCompPars} among 7,204 \tess\ Objects of Interest \citep[TOIs;][accessed on 1 August 2024]{TOI_catalogue} out of which 5,068 are currently flagged as planet candidates \citep[PCs;][accessed on 1 August 2024]{ExoFOP}.  Notable discoveries to date from \tess\ include complex multiplanet systems e.g. TOI-178 \citep{toi178}, TOI-561 \citep[][]{TOI561}, HD 23472 \citep[][]{HD23472_2,HD23472},  habitable-zone super-Earths e.g TOI-700 \citep{TOI700}, TOI-715 \citep{TOI715}, TOI-2095 \citep[][]{toi2095}, circumbinary planets e.g. TOI-1338 \citep{toi1338_1} and TIC 172900988 \citep{TIC172900988}, gas giants orbiting M Dwarfs \citep[e.g.][]{bryant23, toi5205, mdwarf_candidates} and a planet transiting a white dwarf \citep[WD 1856+534\,b;][]{2020Natur.585..363V}. 

\subsection{PLATO}
\label{sec:plato}
Planetary Transits and Oscillations of Stars \citep[\plato;][]{plato_2024} is an ESA M class mission which is planned to be launched towards L2 at the end of 2026. In order to discover terrestrial planets in the habitable zone of Sun-like stars, \plato\ will observe a fixed field in the Southern hemisphere for at least 2 years; this is described in more detail in\,\autoref{sec:lops2}. \plato\ focuses on observing mainly FGK stars within the field. Depending on their location in the field, stars are monitored by either 6, 12, 18 or 24 cameras. In order to optimise the solar irradiation falling on the solar panels \plato\ will rotate by 90° every three months and use this data gap to downlink data. \\
\plato\ has $26$ cameras. $24$ of these, combined in groups of $6$ with a bandpass between 500 and 1000\,nm and observing at a cadence of \SI{25}{\second}, are called normal cameras (NCAMs). The remaining $2$ so called fast cameras (FCAMs) will observe only bright targets ($\sim$ 300 targets of V$<8.5$ mag in the centre of LOPS2) at a cadence of \SI{2.5}{\second}, one with a bandpass between 505 and 700\,nm  and one with a bandpass between 665 and 1000\,nm \citep{plato_2024}.  Each normal camera has four CCDs with detector dimensions of 81.18 mm × 81.18 mm, pixel sizes of 18 $\mu$m x 18 $\mu$m and a pixel scale of 15 arcsec/pixel. Normal cameras have a field of view of 1037 deg$^2$. Due to arranging the cameras in four groups of six, overlapping in the centre of the field, \plato's overall field of view is $\SI{2132}{\deg\squared}$. Due to the larger target sample of the normal cameras, we will only focus on the stars that they will observe in this work. \\
\plato\ will downlink unprocessed imagettes, which are cut-out images of selected stars containing the target and the pixels around it (typically 6$\times$6 pixels) and pre-processed lightcurves every 3 months, which will be further processed on the ground and complemented with ground-based follow-up data. Final data products will be summarised into  catalogues of candidates and confirmed planetary systems \citep{plato_2024}.

\subsection{The PLATO LOPS2 Field}
\label{sec:lops2}
\begin{figure*}
    \centering
    \begin{subfigure}[b]{0.42\textwidth}
      \includegraphics[width=\columnwidth]{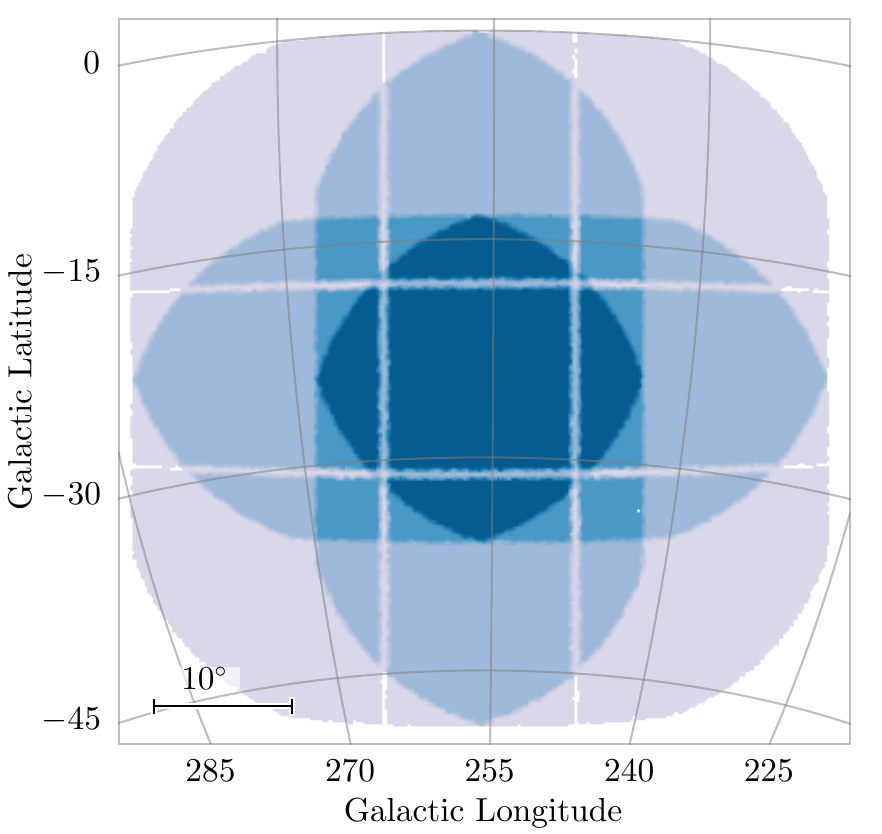}
    \end{subfigure}
    \begin{subfigure}[b]{0.49\textwidth}
      \includegraphics[width=\columnwidth]{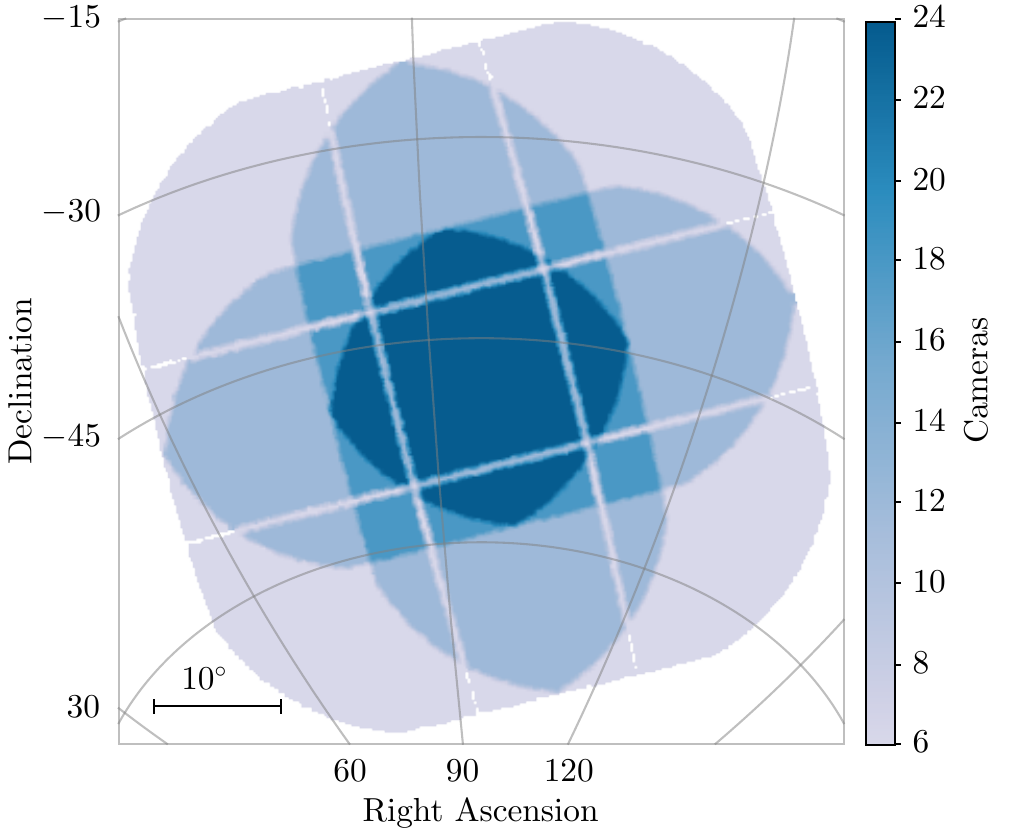}
    \end{subfigure}
    \caption{The PLATO LOPS2 field. Different shades of blue represent the different number of cameras the respective area will be observed with. Left: In galactic coordinates. Right: In equatorial coordinates}
    \label{plato_field}
\end{figure*}

\begin{figure*}
    \centering
    \begin{subfigure}[b]{0.9\textwidth}
      \includegraphics[width=\textwidth]{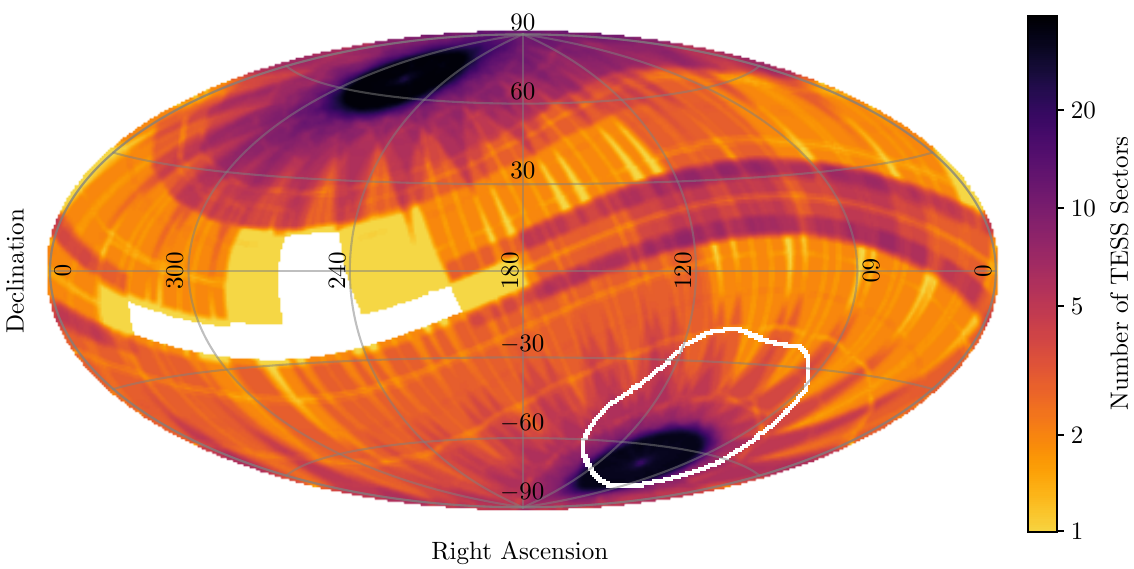}
    \end{subfigure}
    \begin{subfigure}[b]{0.45\textwidth}
      \includegraphics[width=\columnwidth]{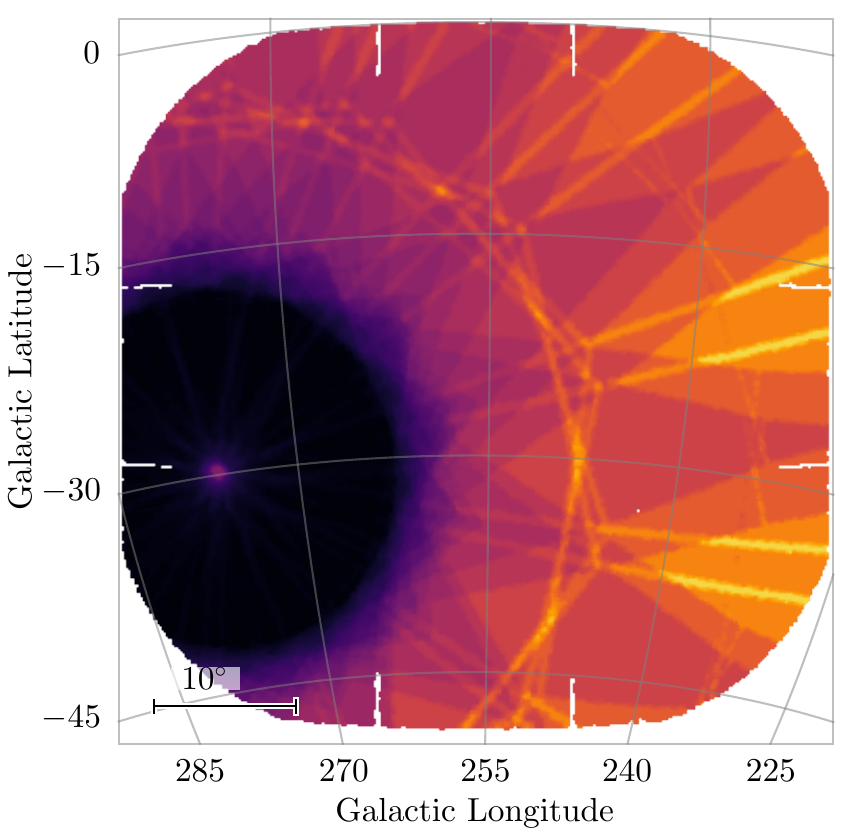}
    \end{subfigure}
    \begin{subfigure}[b]{0.45\textwidth}
      \includegraphics[width=\columnwidth]{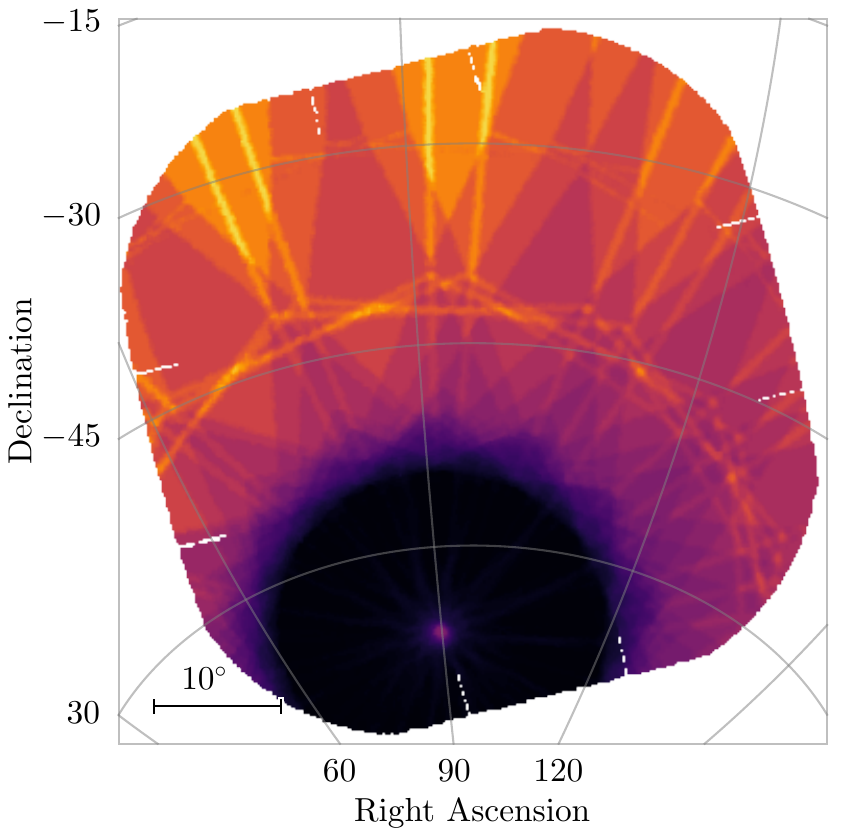}
    \end{subfigure}
    \caption{Overlap of the \plato\ field with \tess's all-sky monitoring. Top: Map showing the number of sectors of \tess\ observing the sky in equatorial coordinates. Dark purple represents the continuous viewing zone, where stars have more than 30 sectors of data. The sky position of the \plato\, LOPS2 field is shown (white line).  The \plato\, LOPS2 field has a large overlap with the \tess\ continuous and near-continuous viewing zones. Bottom Left: Number of \tess\ sectors within the \plato\ field, plotted in galactic coordinates. \tess's continuous viewing zone is coloured in dark purple at the left of the field. Bottom Right: Number of \tess\ sectors within the \plato\ field, plotted in equatorial coordinates. \tess's continuous viewing zone is coloured in dark purple at the bottom of the field.}
    \label{fig:plato_tess_field}
\end{figure*}

In contrast to the \tess\ observing strategy, \plato\, will focus on the single Long-duration Observation Phase South (LOPS2, see \autoref{plato_field}) field in the Southern hemisphere for at least the first two years of the \plato\, mission \citep{plato_2024}.  The LOPS2 field, and its selection process, is described in full in \citet{plato_field_pertenais} and \citet{platofield}. LOPS2 is centered at RA= $6^\text{h}21^\text{m}14.5^\text{s}$ and Dec=$-47\degree53^{'}13^{''}$ (l=255.9375\,deg and  b=-24.62432\,deg in galactic coordinates; \citealt{plato_2024}), corresponding to $\sim5\%$ of the sky due to the field of view of \plato. The LOPS2 field overlaps with \tess's southern continuous viewing zone (see \autoref{fig:plato_tess_field}).  The LOPS2 field runs all the way from the galactic plane (galactic latitude $b$=-0.25\,°) down to a galactic latitude of $b$=-49\,°.  The centre of the field is observed by $24$ cameras, the corners of the field will be observed by 6 cameras, and intermediate overlap zones are covered by either 12 or 18 cameras. 

\subsection{The PLATO Input Catalog}
\label{sec:PIC}
The all-sky \plato\, Input Catalogue \citep[PIC;][]{PIC} consists of 2,675,539 stars, and the stellar parameters are derived from the Gaia survey \citep{gaia, gaia_dr3}. 
The PIC is divided into four samples (P1, P2, P4, and P5) based on different criteria \citep{PIC, plato_2024}. P1 and P2 contain the brightest and most quiet FGK dwarfs and subgiants (V$\leq$11 mag for P1 and V$\leq$8.5 mag for P2; with estimated noise of $\leq$ 50 ppm/h). P4 contains M dwarfs brighter than V=16\,mag, and P5 is a statistical sample covering FGK dwarfs and subgiants brighter than V=13\,mag.  Hereafter we will refer to samples of stars within the PIC (and each PIC subset) in the LOPS2 field by adding the prefix "LOPS2" (e.g. "LOPS2 PIC", "LOPS2 P1").
The LOPS2 PIC (v.2.0.0) contains $179,566$ stars. The number of stars in each of the four PIC samples in LOPS2 are shown in \autoref{tab:pic_samples}.
For the P1, P2 and P4 samples data will be available in the form of imagettes monitored at a cadence of 25 seconds, while stars in the P5 sample will have a mix of imagettes and light curves with cadences of 25, 50 or 600 seconds depending on each star \citep{plato_2024}.\\
Relying on Gaia DR3 data \citep{gaia_dr3}, the PIC catalogues parameters such as magnitude, radius, mass, effective temperature of stars of interest to \plato.  The PIC also contains the Gaia DR3 flags for non single stars, dividing them between photometric, spectroscopic and astrometric binaries.   The expected number of \plato\ cameras observing each star, as well as the expected systematic and random noise for each star, is recorded in the PIC \citep{boerner2024plato_noise}. \\

\section{Methodology}
\label{sec:method}
\subsection{TESS Monitoring of PLATO LOPS2 field}
\begin{figure}
    \centering
    \includegraphics[width=\columnwidth]{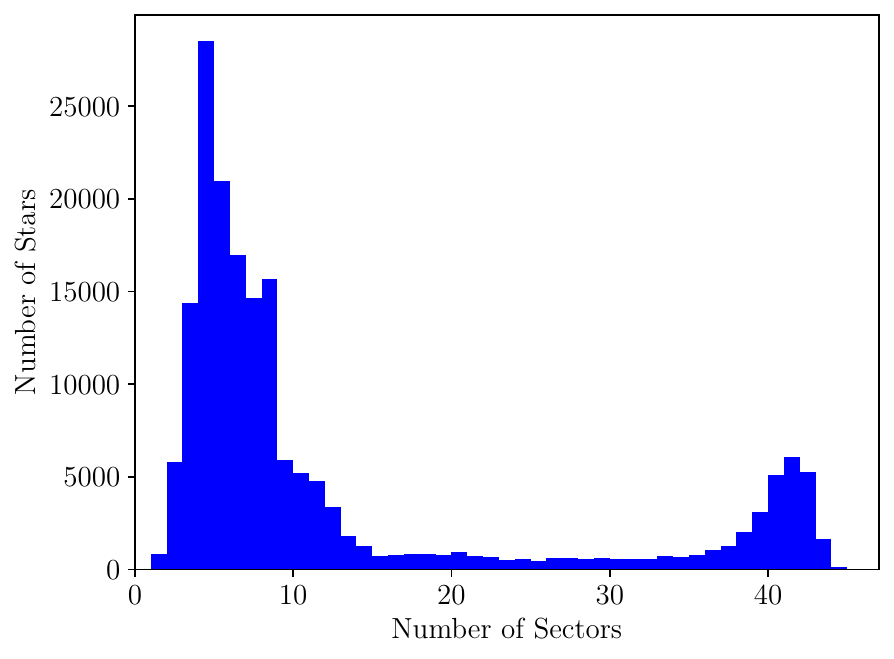}
    \caption{Histogram showing the number of \tess\ Sectors observed for each of the 179,566 Target PIC stars in the LOPS2 field based on Years 1-7 of the \tess\ mission.  The peak near 40 Sectors is due to stars in the \tess\ southern CVZ. } 
    \label{fig:sector_pipline}
\end{figure}

We use TOPCAT \citep{TOPCAT}, an interactive graphical viewer used to edit tabular data, to cross-match the LOPS2 PIC with the target lists of the Science Processing Operations Centre \citep[SPOC;][]{spoc} and the Quick Look Pipeline \citep[QLP;][]{qlp, qlp_2, qlp3, qlp4} on Exact Values using their TIC IDs. We find that the majority of the stars will be monitored in \plato's LOPS2  PIC already have \tess\ lightcurves created by SPOC and QLP. \\
We find that bright stars within the LOPS2 PIC (the LOPS2 P1 and LOPS2 P2 sample) are covered nearly completely by SPOC and QLP. In the statistical sample (LOPS2 P5) $\sim$70\% of the stars are covered by SPOC due to its limit in producing no more than 160,000 FFI light curves per sector, while QLP covers LOPS2 P5 nearly completely. The biggest difference between the two pipelines is found in the coverage of the M dwarf sample (LOPS2 P4). Due to SPOC's selection function's cutoff, only $\sim$30\% of P4 are covered, while QLP covers more than 85\%. However due to the P4 sample being as faint as V=16\,mag several stars are also not covered by QLP.   See \autoref{tab:pic_samples} for the breakdown of the \tess\ data products for each of the LOPS2 PIC samples.  We are hence able to analyse P1 and P2 in more detail by using SPOC data. P1 contains $\sim$60\% F dwarfs and subgiants, $\sim$30\% G dwarfs and subgiants, $\sim$10\% K dwarfs and subgiants\\
So far \tess\ has observed the \plato\, field in three of its years. Using TESS-Point \citep{tesspoint}, we determined how many sectors each star in the PIC was monitored by \tess\ up to year 7 of the mission as shown in \autoref{fig:sector_pipline}. The \plato\, field partially overlaps with \tess's continuous viewing zone and hence a peak of stars being observed in more than 40 sectors can be found. For these stars more than 1,000 days of data are available. Based on this bimodal distribution we divide the LOPS2 PIC stars into two samples; one with $\leq$20 sectors of \tess\, monitoring which we will refer to as non-continuous viewing zone (non-CVZ) and stars with $>$20 sectors of \tess\, which we will refer to as near or within the continuous viewing zone (CVZ).  We note that approximately 80\% of the LOPS2 PIC stars are in this non-CVZ sample, while 20\% are in the CVZ sample.

\begin{table*}
    \centering
    \begin{tabular}{c|ccccccc}
        \hline
        Sample & Description &VMag & Noise & Type &  Number & Stars in & Stars in \\
         &  & & (ppm/h) & & of Stars& SPOC LCs & QLP LCs \\
        \hline 
        P1 & bright FGK dwarfs and subgiants & $\leq$13 & $\leq 50$ & F5-K7 & 9,552 & 9,244 & 9,544  \\
        P2 & very bright FGK dwarfs and subgiants & $\leq 8.5$ & $\leq 50$ & F5-K7&  699 & 684 & 699 \\
        P4 & M dwarfs &$\leq 16$ & & M  & 12,414 & 3,991 & 10,757 \\
        P5 & statistical sample  of FGK dwarfs and subgiants&$\leq 13$ & & F5-K7 & 167,152 & 115,960 & 167,052 \\
        \hline
    \end{tabular}
    \caption{Number of stars within each PIC sample and their coverage by SPOC and QLP. }
    \label{tab:pic_samples}
\end{table*}

\subsection{Photometric Precision of TESS and PLATO}
\label{sec:cddp}

\begin{figure}
    \centering
    \includegraphics[width=\columnwidth]{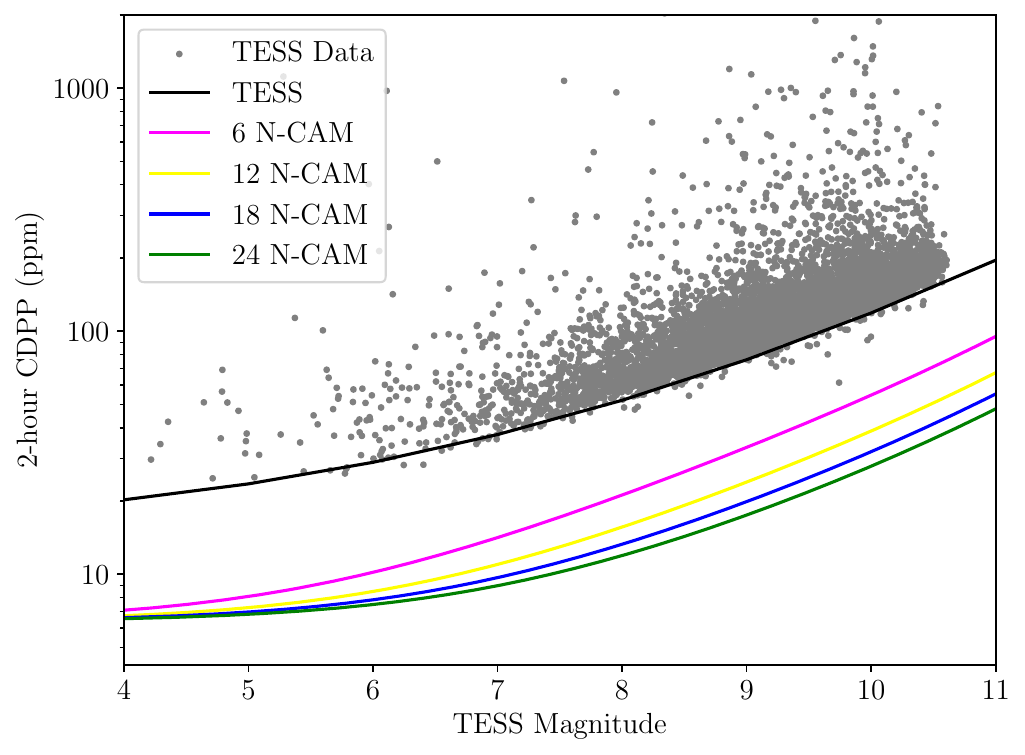}
    \caption{The 2-hour photometric precision as a function of $T$ magnitude.  \tess\ 2h-CDPP precision for the 9,244 stars in the \plato\, P1 sample in the LOPS2 field covered by SPOC (grey points).  Overplotted is the noise model fit of the \tess\ CDPP (black line) and the estimated \plato\ precision \citep{boerner2024plato_noise} for 6, 12, 18, and 24 cameras (pink, yellow, blue and green lines respectively)}.
    \label{fig:cdpp}
\end{figure}

\begin{figure}
    \centering
    \includegraphics[width=\columnwidth]{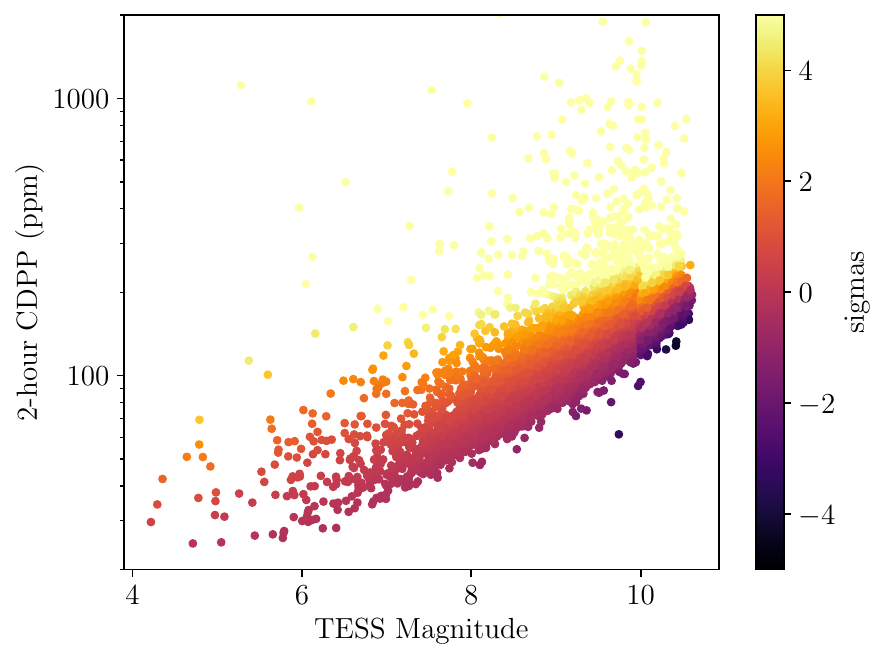}
    \caption{\tess\ 2h-CDPP precision for the 9244 stars in the \plato\ P1 sample. Colours show the number of sigmas each star is away from the CDPP mean of the respective magnitude bin. }
    \label{fig:cdpp_sigmas}
\end{figure}
Variability in light curves can be of instrumental or astrophysical origin. Both can impact transit detections. While instrumental noise is dependent on the telescope, astrophysical variability could arise from high variability and eclipsing binaries. 
The Combined Differential Photometric Precision \citep[CDPP;][]{kepler_cdpp}, introduced during the \kepler\ mission, is a metric that quantifies the photometric variability in a light curve over a particular timescale.  \\ 
For \tess\ lightcurves, the 2-hour CDPP is calculated by the SPOC pipeline and made available in the header of the fits file (keyword: \texttt{CDPP2\_0}). The CDPP was first calculated for the Kepler pipeline and is defined as the root mean square (RMS) photometric noise on transit timescales \citep{kepler_cdpp}, which is 2 hours for \tess. QLP data products do not contain a photometric precision metric. In \autoref{fig:cdpp} we plot the 2-hour CDPP for all of the 9,244 \plato\ P1 stars in the LOPS2 field. 
%We also plot the best-fitting 50th percentile CDPP model based on the empirical fit of all 230,000 SPOC 2 minute light curves from \citep{2022AJ....163..290K} and  scale it to 2 hours.  
%We note that the P1 stars in the LOPS2 field have photometric precision similar to the model that is derived from the full sky, indicating that the photometric precision of the \tess\ data in the LOPS2 field is not unusual compared with the mean \tess\ precision.  \\
As \plato\ is yet to launch we rely on predictions for the photometric noise.  We use the prediction lines from the \plato\ Instrument Noise Estimator \citep[PINE;][]{boerner2024plato_noise}.  The noise estimate curves from PINE are presented in terms of the $V$ band magnitude of the stars.  In order to compare directly to the \tess\ data and curve \citep{2022AJ....163..290K}, we convert these to $T$ band magnitude by assuming a colour term of ($V$-$T$=0.6), typical for solar-like stars in the \plato\ P1 sample.  PINE also presents photometric noise on a 1-hour timescale.  To compare with \tess\ data we convert the PINE noise estimates to a 2-hour timescale by assuming white noise and dividing by $\sqrt{2}$.\\
 Since the \plato\ noise curves do not include noise arising from stellar activity, we fit a model to the minimum CDPP of the TESS data. We use the noise model by \cite{2022AJ....163..290K} and fit it to the lowest CDPP value in each magnitude bin after removing outliers with a 5 iteration 2.5 $\sigma$ clip. This is shown by the black line in \autoref{fig:cdpp}.
We compare the measured \tess\ photometric precision to the estimated \plato\ precision for P1 stars in LOPS2 in \autoref{fig:cdpp}. 
We plot four different scenarios for \plato\ based on whether the star is observed in 6, 12, 18, or 24 \plato\ cameras (see \autoref{sec:plato}).

\subsection{High Photometric Variability Stars}
\label{sec:noisy_stars}
In order to identify stars in the \plato\ P1 sample in the LOPS2 field that may have excess photometric astrophysical noise (caused by binary variability, pulsations, stellar activity variability), we calculate the excess photometric noise in the \tess\ light curve.  To do this we divide the P1 sample into bins of one magnitude in width. We calculate the mean and standard deviation of the CDPP distribution in each bin ignoring CDPP values that are above the 90th quantile of the distribution. For each star within this bin we calculate its CDPP offset from the bin's mean and record this value in multiples of the bin's standard deviation (see \autoref{fig:cdpp_sigmas}).

\subsection{Sensitivity Maps}
\label{sec:sensitivity}
\begin{figure*}
    \centering
    \includegraphics[width=\textwidth]{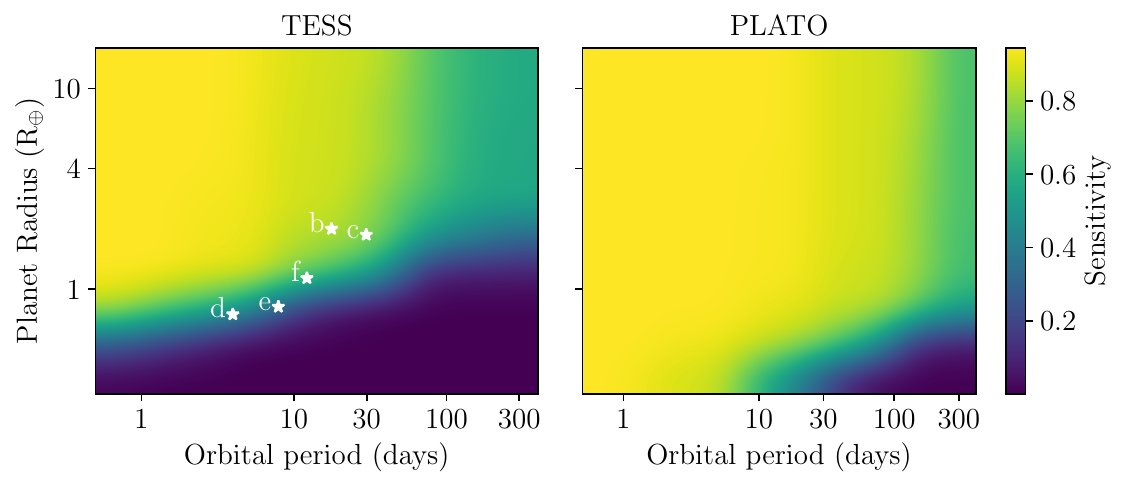}
    \caption{Sensitivity maps of the multiplanetary system HD 23472 (VMag=9.73, R$_S$=0.7\,R$_\odot$, Teff=4684\,K from \citet{HD23472} and a minimum 2-hour CDPP value of 84.8 ppm$/$h in sector 3 of TESS observations).  The positions of HD 23472\,b, c, d, e and f are plotted in white.  \textit{Left}:  The sensitivity for the twelve sectors (Sectors 1, 2, 3, 4, 11, 29, 30, 31, 34, 64, 68, 69 observed; sectors 95 and 96 to be observed within the next year) observed in Year 1, Year 3 and Year 5 of the \tess\ mission.  \textit{Right}:  Simulated sensitivity for \plato\, data assuming two years of \plato\, data with gaps of 24\,hours per quarter and \texttt{BOLrandomNSR} noise value from the PIC. }
    \label{fig:HD23472}
\end{figure*}

\begin{figure*}
    \centering
    \includegraphics[width=\textwidth]{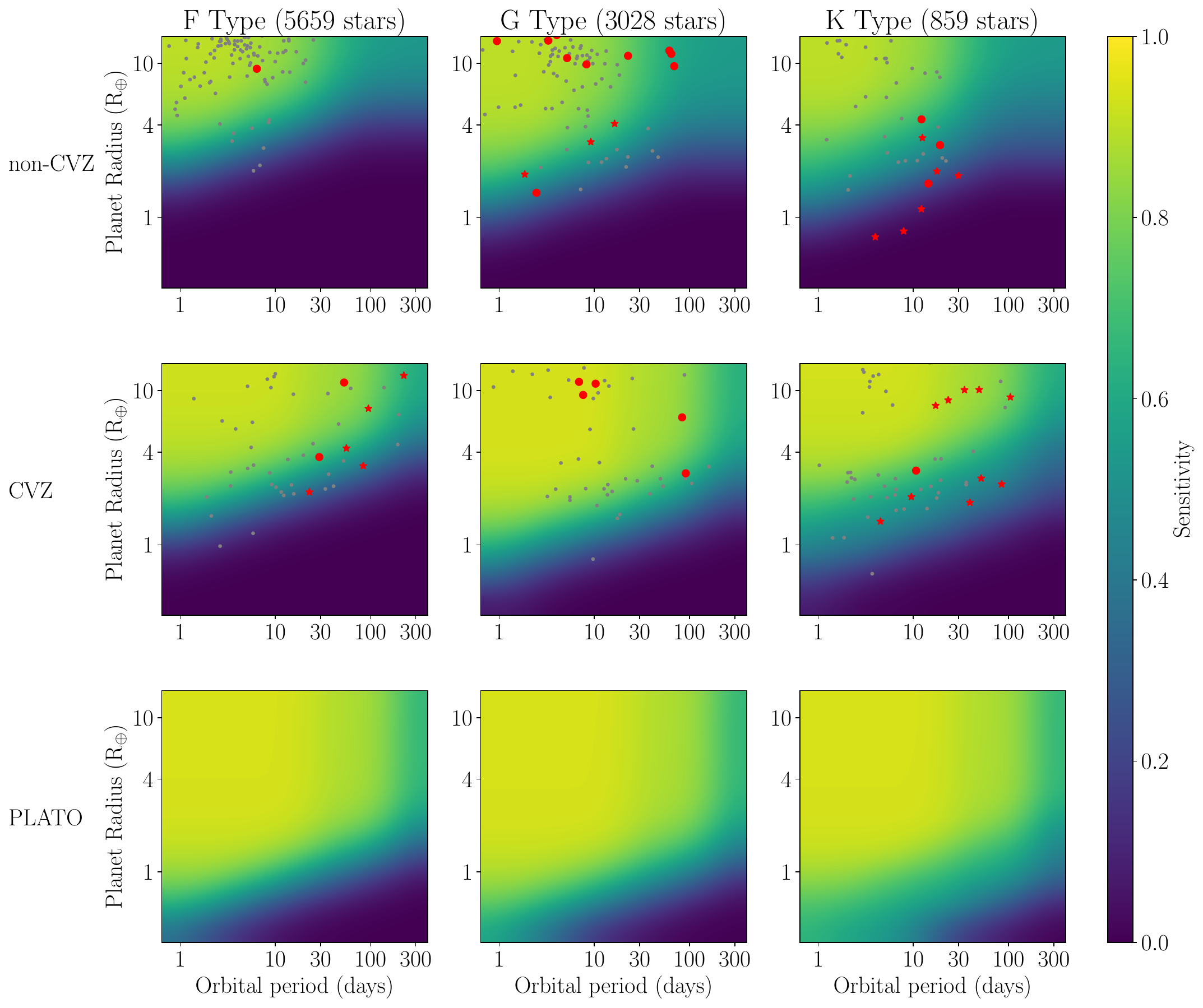}
    \caption{Combined sensitivity maps for the FGK dwarfs of the P1 LOPS2 sample. Yellow and light green regions show where planets can get detected with a high sensitivity, while dark blue regions show areas where the respective telescope is not sensitive to detect planets. Confirmed planets detected by \tess\ in the LOPS2 field are plotted in red, while multiplanetary systems with at least one transiting planet found by \tess\ are marked with a red star. Planet candidates identified by \tess\ in the LOPS2 field are plotted in gray. Top: Sensitivity for stars in the \tess\ non-continuous viewing zone ($\leq$20 sectors). Middle: Sensitivity for stars in the \tess\ continuous viewing zone ($>$20 sectors). Bottom: Expected sensitivity for \plato. Left: $\sim$60\% F dwarfs and subgiants. Middle: $\sim$30\% G dwarfs and subgiants. Right: $\sim$10\% K dwarfs and subgiants.}
    \label{fig:sensitivities}
\end{figure*}

To explore the parameter space where \plato\, will have the greatest potential for new discoveries of transiting exoplanets, we create sensitivity maps for \tess\, and \plato.  These maps are set out in terms of planetary radius and orbital period space.  They are a function of each survey's photometric precision and duration of monitoring.  It is important to stress that they show only the sensitivity of the survey to transiting planets, and are agnostic as to the occurrence rate of planets at any particular radius or orbital period.\\
To calculate the sensitivity maps for \tess\ and \plato\ we use the Transit Investigation and Recovery Application \citep[TIaRA;][]{TIARA}.  TIaRA works on a star by star basis. It reads in the timestamps of each star, the photometric noise in terms of a CDPP value, and the stellar radius which is obtained from the PIC.  It then calculates the signal-to-noise for a large number of exoplanet transits at randomly generated radii, orbital periods, impact parameters and phases.  Exoplanet transits are deemed detected with a probability based on the signal-to-noise of the transit and a gamma-function selected based on the number of transit events in the data.  The details of TIaRA are fully described in \citet{TIARA}.\\
To calculate sensitivity maps for \tess, we apply TIaRA to each star in LOPS2 P1.  TIaRA reads in the available SPOC FFI lightcurve for each sector in which the star was monitored by \tess\ and extracts the timestamps with good photometric data (quality flags = 0).  From the FITS header of the SPOC lightcurve, TIaRA reads in the 2-hour CDPP noise, the crowding metric, the stellar radius, the effective temperature and the \tess\ magnitude.\\
We modified TIaRA to generate 1,000 transiting planets per star uniformly distributed in $\log_2$ space following occurrence rate studies of \cite{hsu_occurence_rates} for periods between 0.5 and 400 days and radii between 0.3 and 16 R$_\oplus$. We also modified TIaRA to no longer perform a minimum detectable radius cut-off.  Since we are interested here in sensitivities rather than yields, we did not apply the TIaRA functions for including the geometric probability of transit or the planet occurrence rate. 
For the \tess\ sensitivity maps, we only use the available data. Hence the sensitivities will improve with more \tess\ observations of each star.\\
Although we do not yet have real \plato\ light curves, we can simulate \plato\ data based on the estimated noise performance by PINE \citep{PLATOnoise, boerner2024plato_noise} and the plans for the first two years of the \plato\ mission \citep{plato_2024}. Other \plato\ tools like the PLATO Solar-like Light-curve Simulator \citep[PSLS;][]{PSLS} or PlatoSim \citep{platosim} also provide similar simulations. We use the stellar parameters (radius, mass, temperature) recorded in the PIC \citep{PIC}. The PIC also records an estimate for the beginning-of-life random and systematic noise \citep[BOLrandomsysNSR;][]{boerner2024plato_noise}. This noise value takes into account how many \plato\ cameras will observe each star. We generate timestamps assuming an observation duration of two years, with a cadence of 25\,s and one day data gaps every three months for spacecraft rotation and data downlink.  This last assumption is based on the operation of the Kepler spacecraft, which also performed a single-field stare from a sun-centred orbit \citep{kepler} and had a downlink time of $\sim$0.9 days on average \citep{kepler_data_gap}.  We bin the timestamps into bins of 10 minutes, and scale noise accordingly, for computational efficiency.  We then run TIaRA with these timestamps and noise properties in the same manner as we ran it for the \tess\ data.\\
TIaRA outputs detection probabilities for \tess\ and \plato\ for transiting planets over a range of orbital periods and radii for each star in the LOPS2 P1 sample. We bin this data into 12 bins in radius and 10 bins in orbital period. From this data we create the sensitivity plots for each star in the LOPS2 P1 sample.
Examples of our  sensitivity maps for P1 stars in the \plato\ LOPS2 field are set out in \autoref{fig:HD23472} and \autoref{fig:sensitivities}.

\subsection{Known Systems}
\label{sec:known-systems}
In addition to new discoveries, the \plato\ mission will be monitoring known systems that lie within the LOPS2 field.  These include known transiting exoplanet systems, unconfirmed transiting exoplanet candidate systems, and non-transiting exoplanet systems.  Within LOPS2 there are also transiting brown dwarf systems, eclipsing binaries, and nearby white dwarfs that will all be of interest to the astrophysics community.  \\
In order to find confirmed transiting planets in the LOPS2 field we cross-match the PIC to the transiting exoplanets in the NASA Exoplanet Archive \citep{PSCompPars} and TEPCat \citep{TEPCat}. Since the PIC only covers FGK stars brighter than V=13 mag (and M Dwarfs brighter than V=16 mag), we also perform a cone search in TOPCAT to find all remaining transiting exoplanet systems in the NASA Exoplanet Archive and TEPCat that lie within the LOPS2 field of view, irrespective of magnitude or spectral type. \\
We perform a similar cross-match for non-transiting exoplanet systems (\autoref{sec:rv_planets}), transiting brown dwarfs (\autoref{sec:transits}), eclipsing binaries (\autoref{sec:ebs}), and white dwarfs (\autoref{sec:wd}) that lie within the LOPS2 field.

\section{Results and Discussion}
\label{sec:results}
\subsection{Photometric Precision}
\label{sec:precision_results}
For the P1 sample of stars in the LOPS2 field, the \tess\ photometric precision is set out in \autoref{fig:cdpp}.  As expected, the estimated \plato\ photometric precision is significantly better than the \tess\ precision over all magnitudes.  At the bright end ($T$=4), we estimate that the \plato\ light curves should nominally improve on the \tess\ precision by a factor of approximately three, from 20\,ppm to 7\,ppm.  For fainter stars the precision is more dependent on the number of \plato\ cameras that will monitor the stars.  For stars that are monitored by six cameras, the improvement for a $T$=9 magnitude star is 43.1\,ppm, from 76.3\,ppm to 33.2\,ppm. 
However for the stars monitored with 24 cameras, this improvement over \tess\ is 58.8\,ppm, from 76.3\,ppm to 17.5\,ppm.

\subsection{High Photometric Variability Stars}
\label{sec:high_noise}
\begin{figure}
    \centering
    \includegraphics[width=\columnwidth]{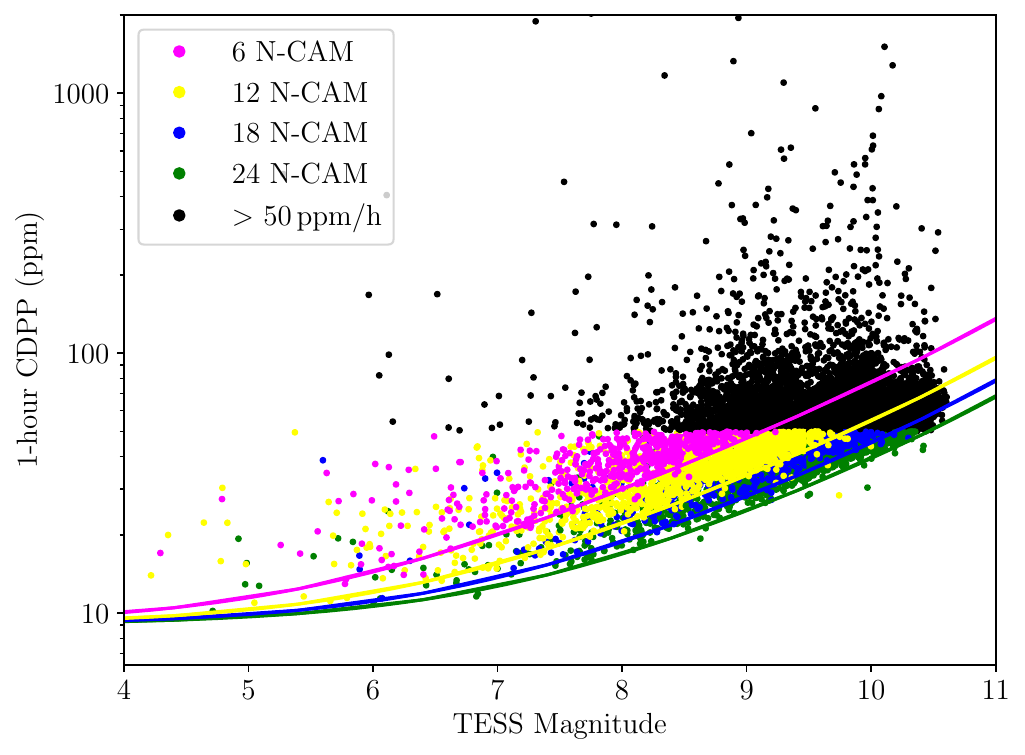}
    \caption{\tess\ CDPP values for the P1 sample scaled to \plato's precision for each camera. Due to the P1 noise cutoff at 50 ppm the different numbers of \plato\ cameras, which are represented by the different colours. have different magnitude cutoffs. Scaled CDPP values above the P1 noise requirement of 50 ppm are coloured black. This affects 40.9\% stars monitored by 24 cameras, 56.9\% for 18 cameras, 60.2\% for 12 cameras and 67.4\% for 6 cameras.}
    \label{fig:scaled_cdpp}
\end{figure}

The \tess\ CDPP values show that while most ($96\%$) of stars within the LOPS2 P1 lie within $5 \sigma$ of the precison distribution (see \autoref{fig:cdpp} and \autoref{fig:cdpp_sigmas}), there are a number of stars that have much higher CDPP noise than we would expect given their magnitudes.  It is important to understand why this is occurring, as detecting transiting planets around these stars will be much more difficult than around photometrically quiet stars.
Some fraction of this photometric noise could be due to systematic noise unique to \tess, such as scattered light from Earth and the Moon or other spacecraft specific noise sources \citep{scattered_light}.  As such this will not be relevant for assessing the likely precision of these stars in \plato.  However, some of the stars with high noise may exhibit true astrophysical variability, in which case we would expect such variability to also be present in the \plato\ data.\\
To investigate this effect, we take the stars with a 5-sigma increase in \tess\ noise from the mean noise within the P1 LOPS2 sample.  These are the stars plotted in yellow in \autoref{fig:cdpp_sigmas}.  We find that SPOC calculates a CDPP value for 9,107 of the 9,244 stars it produces lightcurves for in LOPS2 P1 sample. Out of these, 376 stars  are within this higher 5-sigma noise band.  We inspected these light curves in order to determine the cause of the high CDPP value. We determined that approximately 35\% are due to \tess\ systematic noise, while the remaining 65\% are due to true astrophysical variability. More detailed analysis of stellar variability in \tess\ is beyond the scope of this work but discussed  by e.g. \citet{starclass,fetherolf_stellar_variability}.\\
The definition of the P1 sample has the requirement to contain stars with random noise below 50\,ppm per hour. Using the available \tess\ data and the precisions for \tess\ and \plato, we scale the actual CDPP values of the stars in P1 from \tess\ to \plato\ noise for the different numbers of cameras (see \autoref{fig:scaled_cdpp}) assuming the noise is photometric and there is no systematic astrophysical noise. 
In Kepler data, \cite{kepler_noise} found  more astrophysical noise arising from stars than expected. \plato\ will detect astrophysical noise at a higher precision than \tess\ can, which is not taken into consideration in the \plato\ noise values.  However it may be possible to use the higher precision of \PLATO\ to model and correct for some types of stellar activity. \\As  expected from the \tess\ CDPP in \autoref{fig:cdpp_sigmas}, there are several stars in P1 that have more noise in real data than predicted by PINE. The percentage of stars of higher noise decreases with more cameras. We find that 40.9\% of the stars predicted to be monitored by 24 cameras are above the 50 ppm limit, 56.9\% for 18 cameras, 60.2\% for 12 cameras and 67.4\% for 6 cameras. This noise is caused by astrophysical instrumental effects. These stars are required to be studied further to determine that they meet the 50 ppm noise requirement.
We note the different magnitude cutoff for the different cameras in order to achieve the photometric precision of 50 ppm. 

\subsection{Sensitivities}
\label{sec:sensitivity_results}
In order to quantify the sensitivity to detect transiting exoplanets for each of the LOPS2 P1 stars, we used the sensitivity maps (\autoref{sec:sensitivity}) to determine the smallest planet radius for which we expect to have a 50\% probability of detection for different orbital periods in \tess\ for each star. We select orbital periods of 30 and 100 days and denote these as R\_min,30d and R\_min,100d  respectively as listed in \autoref{tab:PIC}.

\begin{table*}
\begin{tabular}{llllllllll}
\hline
PIC         & TIC       & Teff   & Vmag    & Stellar Radius  & 2-hour CDPP      & \# Sectors  &  R\_min,30d & R\_min,100d         \\
ID         & ID       & (K)   &     & (\rsun)  &(ppm)       &   & (R$_\oplus$) & (R$_\oplus$)         \\
\hline
2761533000026 & 219367750 & 6521.45 & 5.37533 & 1.53946 & 35.27442551 & 5               &       1.937 & 3.55  \\
2787400000022 & 219420836 & 6179.17 & 5.43223 & 2.30214 & 31.41201591 & 6               &       2.68  & 3.82 \\
2809154000107 & 255630992 & 6086.41 & 5.5805  & 2.28453 & 30.98875999 & 11              &       2.05  & 3.58  \\
2784047000097 & 219143616 & 5072.99 & 6.53807 & 2.26656 & 52.7553215  & 7  & 2.97 & 3.93 \\
2768594000018 & 219197061 & 6264.25 & 6.48388 & 2.34305 & 29.8950119  & 8               &       2.12 & 3.69 \\
2809266000285 & 238622063 & 5919.73 & 6.53401 & 2.44669 & 43.26426697 & 8   & 2.86  & 3.85 \\
2785875000326 & 268181496 & 6162.19 & 6.63489 & 1.11451 & 57.92811203 & 7               &       1.89 & 3.50  \\
2802623000303 & 291635915 & 5585.48 & 6.72046 & 1.00965 & 29.67115784 & 8               &       1.615 & 2.09  \\
2768785000027 & 145253043 & 6546.0  & 6.71088 & 1.35983 & 71.34440613 & 6               &       2.19  & 3.68   \\
2467489000191 & 150796339 & 6460.93 & 5.05033 & 1.2725  & 50.95212555 & 4               &       2.16 & 3.69 \\
\hline
\end{tabular}
\caption{\tess\ sensitivities for LOPS2 P1 stars.  Only the first 10 rows are shown here; the full table is available online in machine-readable ASCII form.}
\label{tab:PIC}
\end{table*}

\begin{figure*}
    \centering
    \includegraphics[width=\textwidth]{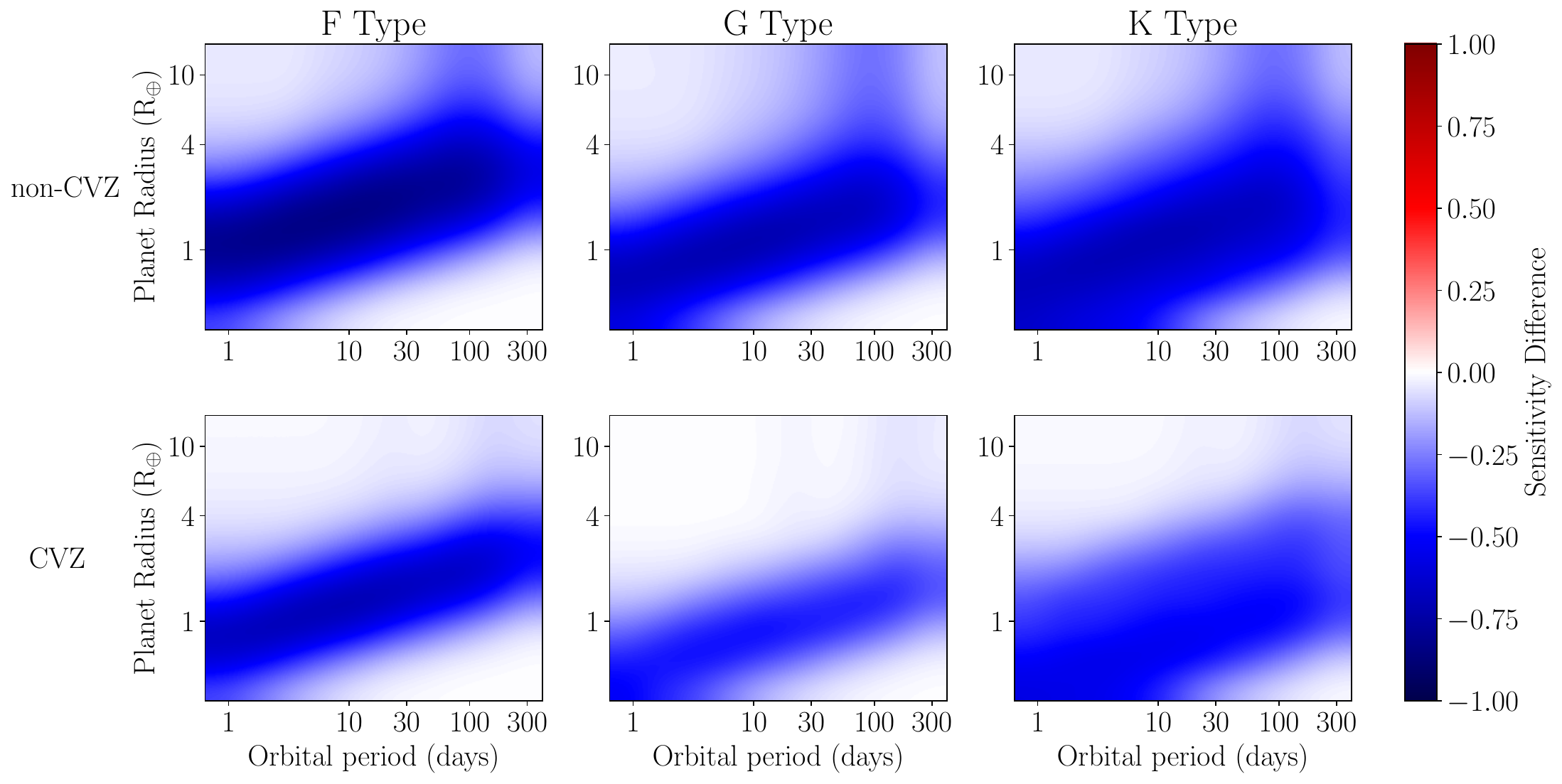}
    \caption{Sensitivity difference between the \tess and \plato\ missions for FGK stars of the P1 LOPS2 sample. Sensitivity difference values $<$0 (blue) show the regions where \plato\ has a higher sensitivity than \tess. Right: F dwarfs. Middle: G dwarfs. Left: K dwarfs. Top:  Sensitivity difference between stars in the \tess\ non-continuous viewing zone ($\leq$ 20 sectors) and \plato\ predictions for all stars of the respective spectral type in P1. Bottom: Sensitivity difference between stars in the \tess\ continuous viewing zone ($>$20 sectors) and \plato\ predictions for all stars of the respective spectral type in P1. }
    \label{fig:sensitivity_difference}
\end{figure*}

Since transit sensitivity is strongly dependent on stellar radius, we split our sample into F, G, and K spectral types. We average all the individual \tess\ and \plato\ sensitivity maps for each spectral type to create a combined transit sensitivity map for populations of stars in the LOPS2 P1 sample to provide an overview of the sensitivities .  We also differentiate between stars in the \tess\ non-CVZ ($\leq$20 sectors), the \tess\ CVZ ($>$20 sectors).  The results are shown in \autoref{fig:sensitivities}.\\
The averaged sensitivity maps for \tess\ and \plato\ for each spectral type give insights as to where the discovery potential of \plato\ lies. Confirmed planets and planet candidates to date lie within the expected discovery space of \tess\ as shown in \autoref{fig:sensitivities}. \plato's sensitivity predictions open up a new discovery space for smaller planets than the ones detected by \tess\ around FGK dwarfs and subgiants, down to 1 R$_\oplus$ and for planets of longer period up to 400 days due to its higher precision and its continuous observations for 2 years.\\
From a qualitative comparison, these \plato\ sensitivity maps are not inconsistent with the expected planet yields by \citet{heller_yield, PLATOnoise} and Cabrera et al. (in prep.).\\
In order to highlight the discovery space that \plato\ will explore, beyond what \tess\ has already reached, we map out the sensitivity differences between the two missions in \autoref{fig:sensitivity_difference}. The sensitivity difference is computed by subtracting the averaged \plato\ sensitivity from the averaged \tess\ sensitivity from \autoref{fig:sensitivities}. Hence the sensitivity difference rages from -1 (blue in the plots) where \plato\ is most sensitive to finding planets in comparison to \tess, to 1 (red in the plots) where \tess\ is most sensitive. 0 sensitivity difference (white in the plots) shows where both telescopes have similar sensitivities.  %The bluest regions of these plots show the areas which \plato\ is most sensitive to finding planets in comparison to \tess. 
We can see there are two major regions where \plato\ will have greater sensitivity than \tess. \\
Firstly, for stars of \tess's non-CVZ, \plato\ is predicted to find planets smaller than those \tess\ is typically sensitive to (below 2-4 R$_\oplus$ depending on spectral type) as well as planets of periods longer than 30 days. For stars in the \tess\ CVZ, planets of longer periods are already well covered by \tess\ alone. Here \plato's most significant contribution will be in discovering planets smaller than 2 R$_\oplus$ that are difficult for \tess\ to detect. \\
\autoref{fig:sensitivity_difference} also shows the radius-period space that cannot be covered by \tess\ or \plato. This covers radii of $<$1 R$_\oplus$ and periods longer than 100 days. \\ 

\begin{figure*}
    \centering
    \includegraphics[width=\textwidth]{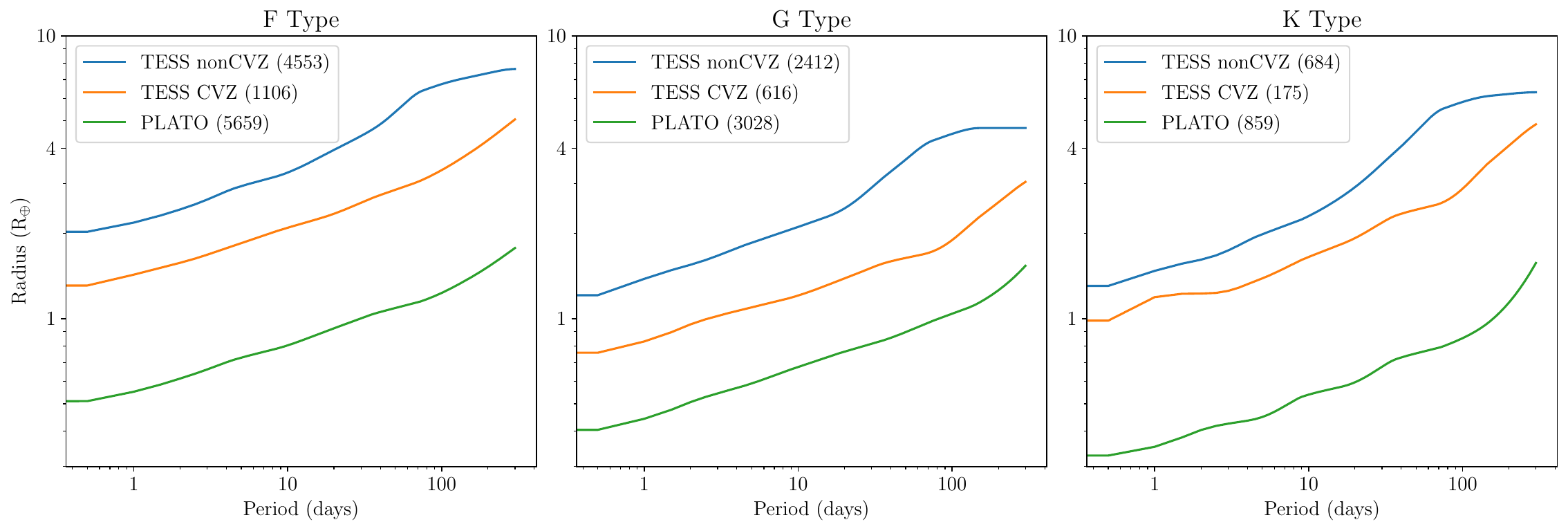}
    \caption{50\% transit sensitivities for F (left) G (middle) and K (right) dwarfs in the P1 LOPS2 sample. 50\% transit sensitivities are shown for \tess\ in the non-CVZ (blue), \tess\ in the CVZ (orange) and \plato\ (green). The number of stars of each sample are shown in brackets.}
    \label{fig:interpolation05}
\end{figure*}

From the combined transit sensitivity maps we fit a spline curve to calculate the threshold at which we expect a 50\% transit detection at a given planetary radius and orbital period.  We again calculate this 50\% transit sensitivity for F, G, and K spectral types.  The 50\% transit sensitivities are calculated for the \tess\ lightcurves (CVZ and non-CVZ) and for the expected \plato\ lightcurves.  The results are shown in \autoref{fig:interpolation05}.

The interpolation curves show the decrease in detectable radii between the \tess\ non-continuous viewing zone, the \tess\ continuous viewing zone, and \plato\ for the three different spectral types. We find the space between the \tess\ non-continuous zone, the \tess\ CVZ and \plato\ to align with the blue regions in the difference plots (see \autoref{fig:sensitivity_difference}). The interpolation curves don't show a significant difference in long orbital periods which is found in the difference plots. This is due to \tess's sensitivity being around 0.5 for long period planets due to monotransits. Although \plato\ will be doing better for long period planets as shown in the difference plots, \tess\ is still able to detect some planets of longer periods as shown by TOI-4562 \citep{toi4562}, a gas giant planet with an orbital period of 225 days, in \autoref{fig:sensitivities} in the LOPS2 field, and even a planet with an orbital period of 483 days, which does not lie in \plato's LOPS2 field  and had 20 sectors of \tess\ data \citep[TOI-4600\,c][]{toi4600}.\\ 
The K dwarf sample in LOPS2 P1 is very small with only 859 stars, hence all three interpolation curves and sensitivity maps are not representative of the larger K dwarf sample that is contained in P5. This is especially the case for the continuous viewing zone which only includes of 175 K dwarfs and subgiants in P1. In order to obtain a more representative sensitivity curve for K dwarfs, we would require a larger sample and could include the K dwarfs from the P5 sample. \\
Some multiplanetary systems (e.g. HD 23472) are detected although the planets lie very close to \tess's detection limits. Since this is a K dwarf system where the sensitivity map and interpolation is not highly representative as discussed above, we look at the sensitivity of the star individually and find that the planets lie right at the boundary of the detectable space (see \autoref{fig:HD23472}). The confirmed multiplanetary systems may contain further transiting planets that are beyond the detection limits of \tess\ and have great potential to be detected by \plato. The same applies to planets that have so far only been measured by radial velocity but not by transit. In particular, the low-mass and small-radius or long orbital period planets might be beyond \tess's detection limit but within \plato's.\\
Knowing this region where \plato\ is most sensitive to discover new planets will guide ground-based follow-up surveys to achieve precisions required to detect the respective planets.  \PLATO has good sensitivity even below 1\,\rearth, however the occurrence rates for this population of small radius exoplanets is currently unknown. 

\subsection{Known Transiting Planets, Candidates, and Brown Dwarfs}
\label{sec:transits}
\begin{figure}
    \centering
    \includegraphics[width=\columnwidth]{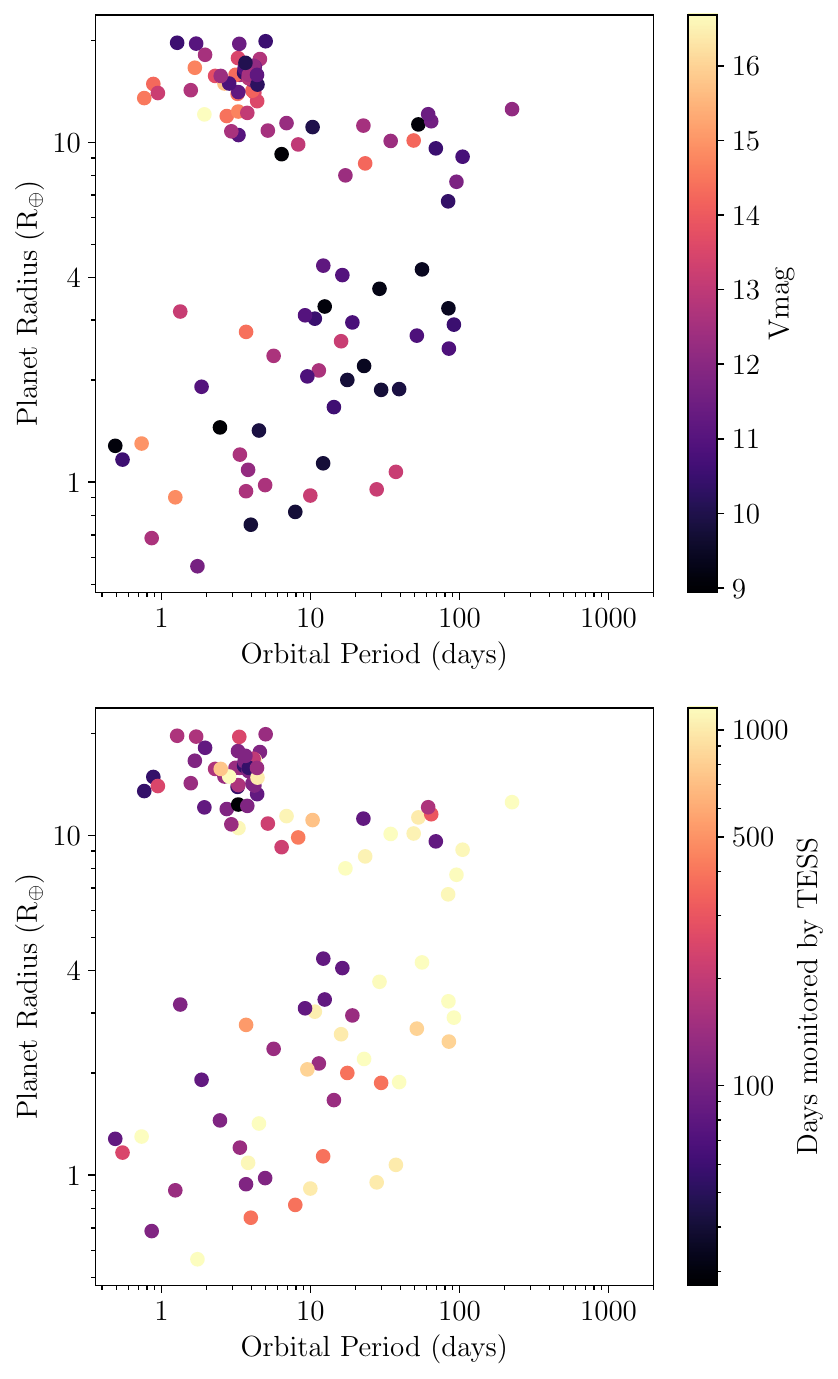}
    \caption{The radius-period distribution of the known transiting planets in the LOPS2 field. Top: Colour-coded to show the V-band apparent magnitude of the transiting planet host star. Bottom: Colour-coded to show the number of days each planet host star has been monitored by \tess.}
    \label{fig:radius_period_planets}
\end{figure}

\begin{figure*}
    \centering
    \includegraphics[width=\textwidth]{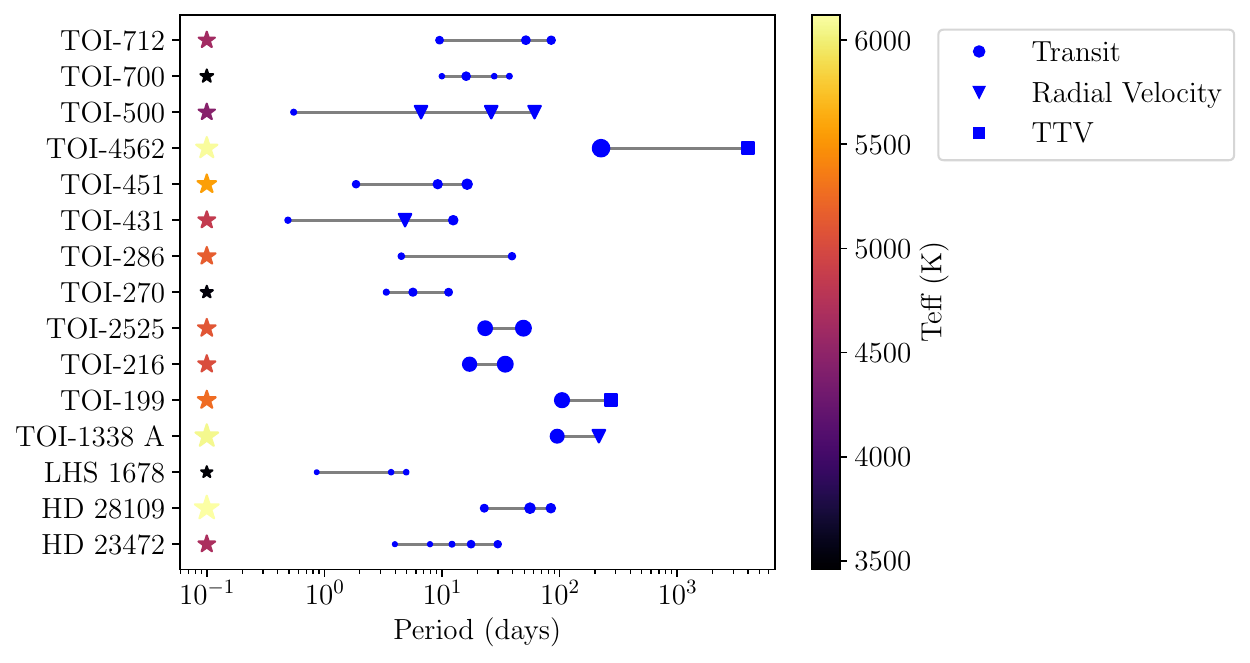}
    \caption{Multiplanetary systems with at least one transiting planet in the LOPS2 field. The size of the star in the plot represents its stellar radius, the colour its effective temperature. Transiting planet are shown in blue circles, the circle size representing their radius. Their distance from the star on the plot corresponds to the planet's orbital period. Systems that also contain a planet that has no transit detection, but by Radial Velocity or TTV are marked by a triangle or square respectively.}
    \label{fig:multis}
\end{figure*}

We find there are currently 101 confirmed transiting planets in the LOPS2 field, orbiting 80 host stars. 65 of these are single-planet systems, and we list these along with the key stellar and planetary properties in \autoref{tab:transit_planets}.  Additionally, 15 systems are multiplanet systems, and these are set out in \autoref{tab:multis}. We note there are 25 transiting planet host stars that are currently not in the PIC since they are either too faint or a different spectral type than FGKM.  A number of the known transiting planet system in LOPS2 were discovered over the past decade by ground-based transit surveys such as WASP \citep[15 systems; ][]{superwasp}, HATS \citep[11 systems; ][]{hatsouth}, NGTS \citep[8 systems; ][]{ngts}, and KELT \citep[2 systems; ][]{Kelt}.  The remaining 44 systems are discoveries from the \tess\ mission, and of these 20 are in the \tess\ CVZ.\\
The radii and orbital periods of the known transiting exoplanets in the LOPS2 field are shown in \autoref{fig:radius_period_planets}. There is a cluster of hot Jupiter type planets with orbital periods between 1-5\,d, and radii between 10-20 \rearth. 
 The remaining planets have radii spanning all the way down from Neptune radii to sub-Earth radii.  We note that the majority of the  currently known transiting planets in the LOPS2 field have periods less than 100 days. Only one transiting planets has a longer period,  TOI-4562 \citep{toi4562}, at 225\,days. We also see that most of the longer period planets (P$>$20\,d) are found around brighter stars that have around 1000\,d of \tess\ monitoring and thus are in or close to \tess's CVZ.\\
There are 15 known multiplanetary transiting systems in the LOPS2 field (see \autoref{tab:multis}).  We present these system in terms of their orbital periods in \autoref{fig:multis}.  Again we can see the majority of these discoveries are at orbital periods $<$100\,d.  Since it is known that many multiplanet systems are highly co-planar \citep{multis_alignment}, \plato\ has the potential for discovering additional transiting planets in these systems, at either longer orbital periods or smaller radii which are challenging to detect in the \tess\ data.\\

\begin{figure*}
    \centering
    \begin{subfigure}[b]{0.45\textwidth}
      \includegraphics[width=\textwidth]{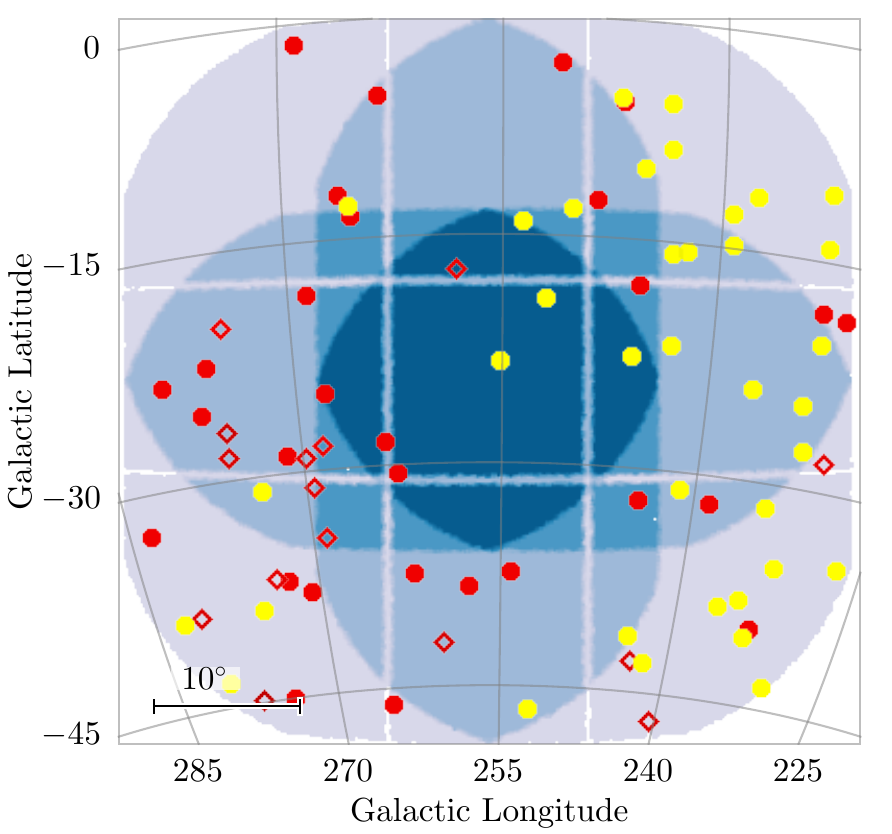}
    \end{subfigure}
    \begin{subfigure}[b]{0.45\textwidth}
      \includegraphics[width=\textwidth]{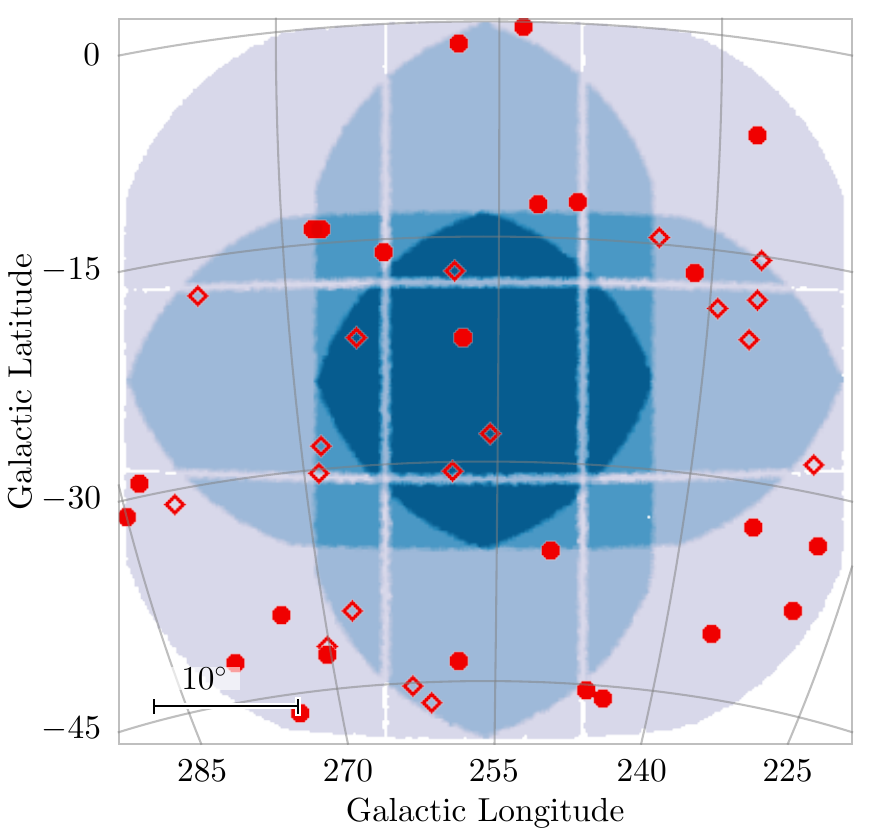}
    \end{subfigure}
    \begin{subfigure}[b]{0.45\textwidth}
      \includegraphics[width=\textwidth]{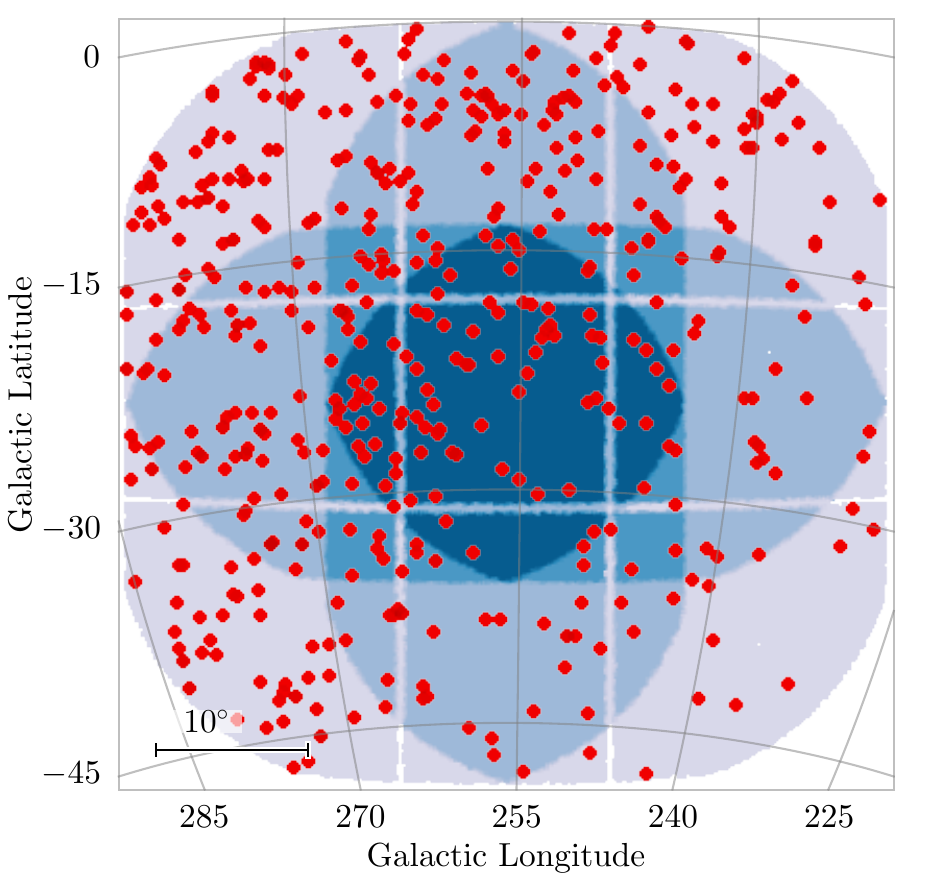}
    \end{subfigure}
    \begin{subfigure}[b]{0.45\textwidth}
      \includegraphics[width=\textwidth]{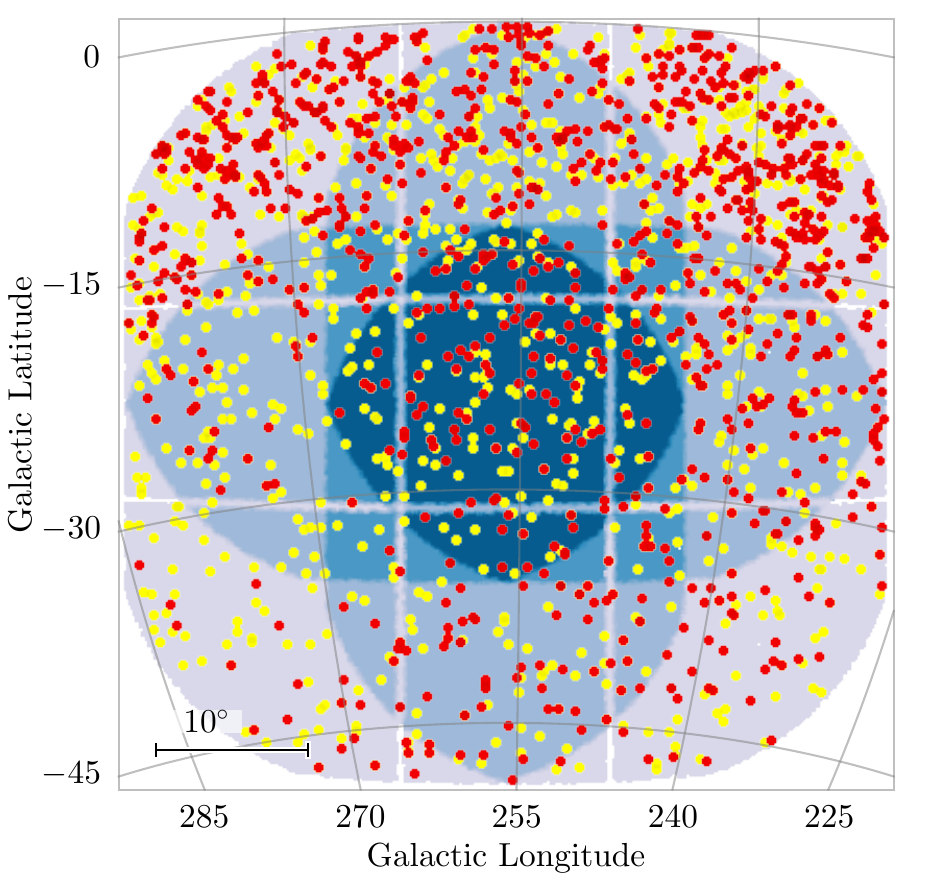}
    \end{subfigure}
    \caption{Top Left: Stars hosting at least one transiting planet in the LOPS2 field. Planets detected by \tess\ are coloured red, Planets detected by other ground-based facilities are yellow. Multiplanetary systems are marked with a diamond, single planet systems with a circle. Top Right: Stars hosting planets detected by radial velocity in the LOPS2 field. Stars with only one detected planet are marked with a circle, stars hosting more than one confirmed planet are marked with a diamond. Bottom Left: \tess\ planet candidates (TFOPWG Disposition 'PC') in the LOPS2 field. Bottom Right: Eclipsing binaries in the PIC. Eclipsing binaries found by \tess\ (Prapotnik Brdnik et al. 2025, in prep) are marked in red, eclipsing binaries flagged by GAIA's NSS flag \citep{2023A&A...674A..16M} (\texttt{NSSflag}=4 in PIC) are marked yellow.}
    \label{fig:plato_fields}
\end{figure*}

We plot the sky distribution of the known transiting planets over the LOPS2 field in \autoref{fig:plato_fields}. We find they are following two distributions. One is that the majority of planets discovered by \tess\ are in or close to \tess's continuous viewing zone. Planets discovered further north are mainly found by other surveys such as NGTS \citep{ngts}, HATS \citep{hatsouth}, KELT \citep{Kelt} and WASP \citep{superwasp}. They are also monitored by \tess\ and are detected by \tess, but most of these planets were already known before the launch of \tess. The planets that are primarily found by instruments other than \tess\ are distributed further to smaller galactic longitudes in the \plato\ field. This is due to the fact that there are not many ground-based observatories at very southerly latitudes. Finally we notice a lack of planets at high galactic longitude and close to $b=0$ in galactic latitude. This is due to the field approaching the galactic plane. The more crowded regions towards the galactic plane cause more contamination for wide-field transit surveys with large pixel scales.  This makes it more challenging to detect transiting planets in these crowded regions. \\
We find $\sim$500 \tess\ transiting planet candidates (PC) in the LOPS2 field by cross-matching from the TOI catalogue \citep{TOI_catalogue} from ExoFOP \citep{ExoFOP} and are flagged as PC by TFOPWG.  We also plot the sky distribution of these \tess\ planet candidates in \autoref{fig:plato_fields}. The planet candidates show a gradient towards the \tess\ CVZ and towards the galactic plane, which is simply due to the higher density of stars in the galactic plane.\\
Many of these \tess\ planet candidates will be confirmed or ruled out before the launch of \plato, although some may remain candidates and new candidates will be added as the \tess\ mission continues. Several planet candidates flagged by TESS are blended due to the large plate-scale of the TESS cameras (21 \arcsec\,pix$^{-1}$). Since PLATO's pixel scale is smaller (15 \arcsec\,pix$^{-1}$), some of these cases may be resolved by PLATO photometry. Currently ruling out blended scenarios is performed with ground-based follow-up (see SG1 \citep{sg1}).\\
There are also a number of known transiting brown dwarfs in the LOPS2 field that we found through manual inspection of the known systems in the LOPS2 field listed on \citet{PSCompPars} and TEPCat \citep{TEPCat}. These typically have masses higher than 13 M$_J$, the deuterium burning limit \citep{Deuterium_burning_limit}, and lower than 80 M$_J$, the hydrogen burning limit \citep{hydrogen_burning_limit}. Due to their similar radii, radial velocity follow-up is required to distinguish transiting brown dwarfs from transiting gas giant planets.  We find there are currently 7 confirmed transiting brown dwarfs in the LOPS2 field, and these are set out in \autoref{tab:brown_dwarfs}. 

\subsection{Known Radial Velocity Only Planets}
\label{sec:rv_planets}
\begin{figure}
    \centering
    \includegraphics[width=\columnwidth]{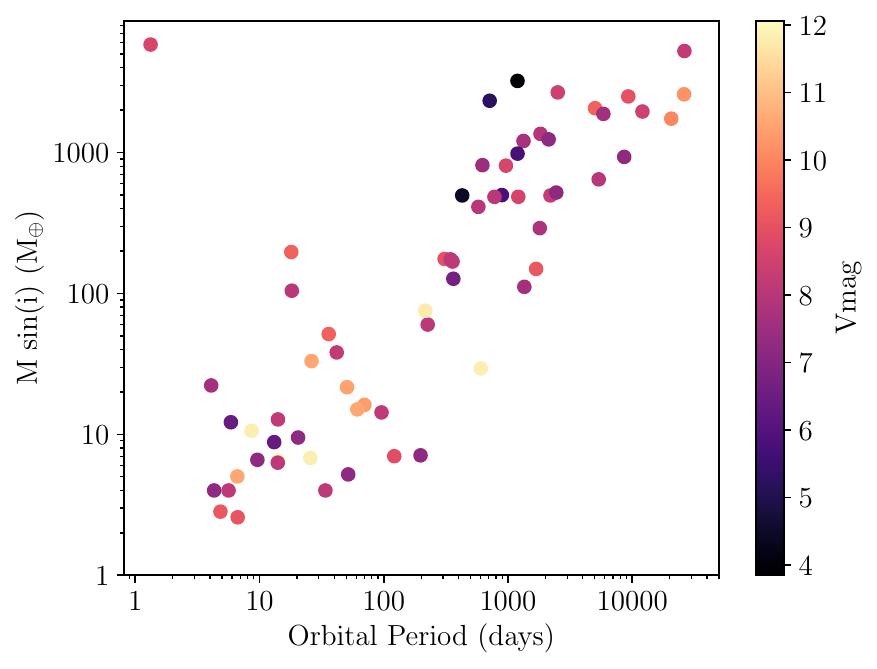}
    \caption{The distribution of minimum mass and orbital period for radial velocity only planets in the LOPS2 field.  The colour scale give the V-band apparent magnitude of the host star.}
    \label{fig:rv_period_radius}
\end{figure}

We also conduct a search for known non-transiting planets (detected by radial velocity alone) in the LOPS2 field using the NASA Exoplanet Archive \citep{PSCompPars}. \\
\plato\ data for these systems could be interesting for many reasons.  The systems  could have additional planets that happen to transit, and hence could be detected by \plato. Several of the radial velocity planets also have periods longer than 100 days and hence a transit geometry may not have been ruled-out yet by \tess\ monitoring. If these planets do transit, \plato\ may be able to detect them. Even if there are no transits in the known radial velocity systems, photometric monitoring with \plato\ will provide valuable  insights into the host stars' stellar parameters and stellar activity.\\
The LOPS2 field contains 43 stars hosting at least one planet that has only been confirmed by radial velocity alone.  24 of these host stars have only one planet, and we detail these in \autoref{tab:rv_planets}.  A further 19 host stars host multiplanetary systems, and these are listed in \autoref{tab:rv_multis}.  We plot the sky distribution of these radial velocity only host stars in \autoref{fig:plato_fields}. Similar to transiting planets, we notice a lack of planets at high galactic longitude and close to $b=0$ in galactic latitude. This is due to the field approaching the galactic plane. The orbital period of these known planets range from just over 1\,d to over 10,000\,d, while the minimum masses range from approximately 3\mearth\ to several \mjup\ (see \autoref{fig:rv_period_radius}).  Due to magnitude limits of radial velocity detections, the majority of radial velocity planets are found around stars brighter than V=12 mag (see \autoref{fig:rv_period_radius}).  Eight radial velocity planet host stars are not included in the LOPS2 PIC sample due to their radius being above the cutoff or falling to close to the edge of the field. 
The 43 confirmed radial velocity systems in the LOPS2 field (see \autoref{tab:rv_planets}) show a much wider distribution of orbital periods than the transiting planets, as seen in \autoref{fig:rv_period_radius}. In fact 27 of these radial velocity systems have planets with orbital periods longer than 100\,days.  Some of these radial velocity planets may not have yet received sufficient photometric monitoring to rule out transits \citep[e.g.][]{transit_rv_planets}, and \plato\ will thus provide useful data for that task. \\

\subsection{Known Eclipsing Binaries}
\label{sec:ebs}
The Gaia Eclipsing Binaries Catalogue \citep{gaia_var2023, 2023A&A...674A..16M} identifies $\sim$ 115,000 eclipsing binary candidates in the \plato\ LOPS2 field based on the Gaia $G$-band stellar light-curves. Most of these are very faint stars, with the distribution peaking at $G$=19 \citep{2023A&A...674A..16M}.  If we limit ourselves to the LOPS2 PIC sample, this is reduced to 1028 eclipsing binary candidates.  \\
The \tess\ $T$-band lightcurves have also been searched for eclipsing binaries. \citet{andrej_EB_paper} presents a catalogue of 4580 eclipsing binary stars found in the first two years (Sectors 1-26) of \tess\ 2-minute cadence SPOC light curves.   
The ephemerides for these eclipsing binaries are estimated using the Quasiperiodic Automated Transit Search  \citep[QATS;][]{qats_1, qats_2}, Eclipse Candidates in Light curves and Inference of Period at a Speedy Rate \citep[ECLIPSR;][]{eclipsr} and box least squares periodogram \citep[BLS;][]{bls} algorithms.  An update to this catalogue is currently in progress \citep[][Prapotnik Brdnik et al. 2025, in prep]{EBs_Kruse}, which will include full-frame image \tess data, hence many more lightcurves.  Accessing this new catalogue (Pr{\v{s}}a, priv comm), we find 1023 eclipsing binaries in the LOPS2 PIC sample.  Cross-matching these with the 1028 LOPS2 PIC eclipsing binaries identified by Gaia, we find an overlap of 229 targets and 794 eclipsing binaries from \tess\ that are $not$ identified by Gaia.  This is not surprising given Gaia lightcurves may be very sparsely sampled - between 16 and 259 photometric measurements \citep{2023A&A...674A..16M} - compared with \tess\ lightcurves which will have thousands of 30-minute candence photometric measurements.\\
We plot the sky distribution of the known eclipsing binaries found by \tess\ and Gaia in \autoref{fig:plato_fields} and find as expected a gradient in distribution towards the galactic plane. This is also analysed in \citet{FP_bray} which notes more false positives to occur towards the galactic plane. \\

\subsection{White Dwarfs}
\label{sec:wd}
Polluted white dwarfs have long hinted that exoplanets orbit white dwarfs \citep{Zuckerman_metal_lines_wd,extrasolar_cosmochemistry, planetary_signatures_wd_wilson}. In more recent years, the detection of transiting planetary debris has provided further evidence to that \citep{vanderburg_transiting_planet_debris, vanderbosch_transiting_planet_debris, Guidry_transiting_planet_debris}. With the launch of \tess\ the first candidate transiting exoplanet was discovered orbiting a white dwarf \citep[WD 1856+534\,b;][]{2020Natur.585..363V}.  Unfortunately, this remains the only transiting exoplanet candidate orbiting a white dwarf and it is not in the LOPS2 field. However, Jupiter-sized exoplanet candidates around white dwarfs have been directly imaged by JWST \citep{2024ApJ...962L..32M}.  With \plato's bluer bandpass and fast cadence, \plato\ could be well-placed to discover further transiting exoplanets around white dwarfs if any are selected for monitoring.  Additionally, photometric observations of white dwarfs are important for studies of pulsating white dwarfs\citep{pulsating_wd} and white dwarf - white dwarf eclipsing binaries \citep{wd_ebs, wd_eb_james} .  Currently, there are no white dwarf stars in the LOPS2 PIC sample as they do not match the spectral type cuts used to produce that sample \citep{PIC}. However these targets could be proposed under the Guest Observer Program \citep[GO;][]{plato_2024}  \\
We perform a cone search of the catalogue of known white dwarfs from GAIA EDR3 \citep{2021MNRAS.508.3877G} brighter than G=13\,mag within the LOPS2 field. We find  RX J0623.2-3741 and $\epsilon$~Ret~B lie in the LOPS2 field. RX J0623.2-3741 is a metal-polluted white dwarf \citep{metal_polluted_wd} and hence an excellent candidate to host planets. $\epsilon$~Ret~B is within a binary system with a K subgiant  \citep[HD 27442][]{white_dwarf_binary_planet} that is hosting a planet (HD 27442\,b) detected by radial velocity \citep{HD27442} as listed in \autoref{tab:rv_planets}. 
White dwarfs would have photometric precision similar to many of the stars in the LOPS2 PIC. Since exoplanet transits around white dwarfs are deeper than around FGK stars, they require less precision, hence also fainter white dwarfs in the LOPS2 field have the potential of exoplanet detection.  

\section{Conclusions}
\label{sec:conclusions}
We have explored the potential for PLATO discoveries in the LOPS2 field, in particular focusing on the P1 sample of bright FGK stars.  We have studied the existing data from the \tess\ mission in order to understand the LOPS2 P1 sample more fully, and to work out the planets that have already been discovered within the LOPS2 field. \\
We find that there are currently 101 known transiting exoplanets in the LOPS2 field, along with $\sim$500 \tess\ planet candidates. Several studies by \citet{heller_yield}, \citet{PLATOnoise}, and Cabrera et. al. (in prep.), which are summarised in \citet{plato_2024}, show an expected yield of up to 5,000 planets in the field.   For all spectral types, \tess\ has very high sensitivity to discovering hot Jupiter type planets, and therefore we expect \plato\ to discover very few if any new transiting hot Jupiters.  For spectral types of F, G and K \tess\ is less sensitive to discovering exoplanets $<$2\,\rearth\ at periods of more than 30 days, whereas \plato\ should have a higher sensitivity to detect such planets.\\
For the 20\% of stars in the LOPS2 P1 sample that lie within the \tess\ CVZ, the \tess\ mission has good sensitivity all the way out to orbital periods of 300\,d.  For these stars, we expect \tess\ will discover the long period giant planets prior to the launch of \plato.  \plato's long period contribution for these stars will therefore be limited to planets with radii $<$4\,\rearth.  However for the remaining 80\% of stars in the LOPS2 P1 sample that are in the \tess\ non-CVZ, we expect \plato\ will discover long period planets down to Earth-size planets which \tess\ was not able to detect due to limited duration monitoring. Combining \tess\ and \plato\ data could also improve sensitivities, especially towards long periods. 
Our work shows the contribution of the \tess\ mission has opened up the long period and small radius space of exoplanets which will be advanced by \plato. 

\section*{Acknowledgements}
The authors gratefully acknowledge the European Space Agency and the \plato\ Mission Consortium, whose outstanding efforts have made these results possible.\\
This paper includes data collected by the \tess\ mission, which are publicly available from the Mikulski Archive for Space Telescopes (MAST). Funding for the \tess\ mission is provided by NASA’s Science Mission directorate. We acknowledge the use of public \tess\ data
from pipelines at the \tess\ Science Office and at the \tess\ Science Processing Operations Center. This research has made use of the Exoplanet Followup Observation Program website, which is operated by the California Institute of Technology, under contract
with the National Aeronautics and Space Administration under the Exoplanet Exploration Program. This research has made use of the NASA Exoplanet Archive, which is operated by the California Institute of Technology, under contract with the National Aeronautics and Space Administration under the Exoplanet Exploration Program. 
IP acknowledges support from the UK's Science and Technology Facilities Council (STFC), grant ST/T000406/1.

\section*{Data Availability}
We make our created catalogue available in a machine readable format.
The light curves used for our CDPP calculations are already available as high-level science products (HLSPs) on MAST, from SPOC. 

% \subsection{Figures and tables}

% Figures and tables should be placed at logical positions in the text. Don't
% worry about the exact layout, which will be handled by the publishers.

% Figures are referred to as e.g. Fig.~\ref{fig:example_figure}, and tables as
% e.g. Table~\ref{tab:example_table}.

% % Example figure
% \begin{figure}
% 	% To include a figure from a file named example.*
% 	% Allowable file formats are eps or ps if compiling using latex
% 	% or pdf, png, jpg if compiling using pdflatex
%     \caption{This is an example figure. Captions appear below each figure.
% 	Give enough detail for the reader to understand what they're looking at,
% 	but leave detailed discussion to the main body of the text.}
%     \label{fig:example_figure}
% \end{figure}

% % Example table
% \begin{table}
% 	\centering
% 	\caption{This is an example table. Captions appear above each table.
% 	Remember to define the quantities, symbols and units used.}
% 	\label{tab:example_table}
% 	\begin{tabular}{lccr} % four columns, alignment for each
% 		\hline
% 		A & B & C & D\\
% 		\hline
% 		1 & 2 & 3 & 4\\
% 		2 & 4 & 6 & 8\\
% 		3 & 5 & 7 & 9\\
% 		\hline
% 	\end{tabular}
% \end{table}

%\section{Conclusions}



\bibliographystyle{mnras}
\bibliography{refs} % if your bibtex file is called example.bib

%\begin{thebibliography}{99}
%\end{thebibliography}

%%%%%%%%%%%%%%%%%%%%%%%%%%%%%%%%%%%%%%%%%%%%%%%%%%

%%%%%%%%%%%%%%%%% APPENDICES %%%%%%%%%%%%%%%%%%%%%

\appendix
\section{Confirmed transiting and radial velocity planetary systems and transiting brown dwarfs in the PLATO LOPS2 field}
\begin{table*}
\begin{tabular}{lllllllllll}
\hline
Host Star & VMag & TMag & R$_S$ & M$_S$ & Teff & R$_P$ & Period & TESS & PLATO & Paper \\
&  &  & (R$_\odot$) & (M$_\odot$) &(K) & (R$_\oplus$) & (days) &  Sectors & Cameras &  \\
\hline
GJ 238 & 11.62 & 9.37 & 0.4 & 0.4 & 3485 & 0.57 & 1.7447 & 42 & 24 & \cite{GJ238}\\
HATS-39    & 12.74 & 12.32 & 1.6  & 1.4  & 6572 & 17.6  & 4.5776   & 4    & 6     & \cite{hats39}            \\
HATS-40    & 13.48 & 12.94 & 2.3  & 1.6  & 6460 & 17.71 & 3.2643   & 4    & -     & \cite{hats39}             \\
HATS-41    & 12.68 & 12.23 & 1.7  & 1.5  & 6424& 14.91 & 4.1936   & 5    & 6     & \cite{hats39}             \\
HATS-42    & 13.68 & 13.15 & 1.5  & 1.3  & 6060& 15.69 & 2.2921   & 6    & -     & \cite{hats39}             \\
HATS-43    & 13.56 & 12.85 & 0.8  & 0.8  & 5099& 13.23 & 4.3888   & 3    & -     & \cite{hats43_44_45_46} \\
HATS-44    & 14.4  & 13.54 & 0.8  & 0.9  & 5080& 11.96 & 2.7439   & 4    & -     & \cite{hats43_44_45_46} \\
HATS-45    & 13.32 & 12.89 & 1.3  & 1.3  & 6450& 14.41 & 4.1876   & 4    & -     & \cite{hats43_44_45_46} \\
HATS-51    & 12.53 & 11.87 & 1.4  & 1.2  & 5758& 15.8  & 3.3489   & 5    & 6     & \cite{hats51}             \\
HATS-55    & 13.53 & 12.93 & 1.1  & 1.2  & 6214& 14.02 & 4.2042   & 4    & -     & \cite{hats54_58}         \\
HATS-66    & 14.28 & 13.64 & 1.8  & 1.4  & 6626& 15.82 & 3.1414   & 5    & -     & \cite{hats60_69}         \\
HATS-76    & 16.68 & 14.86 & 0.6  & 0.7  & 4016& 12.09 & 1.9416   & 3    & -     & \cite{hats74_77}         \\
HD 56414   & 9.22  & 9.1   & 1.8  & 1.9  & 8500& 3.71  & 29.0499  & 41   & -     & \cite{HD56414}           \\
KELT-14    & 11 & 10.39 & 1.5  & 1.2  & 5720& 19.54 & 1.7101   & 6    & 18    & \cite{kelt15}         \\
KELT-15    & 11.39 & 10.68 & 1.8  & 2.1  & 6003& 19.5  & 3.3294   & 9    & 12    & \cite{kelt15}             \\
LHS 1815   & 12.17 & 10.14 & 0.5  & 0.5  & 3643& 1.09  & 3.8143   & 40   & 12    & \cite{LHS1815}            \\
NGTS-1     & 15.67 & 13.91 & 0.6  & 0.6  & 3916& 14.91 & 2.6473   & 5    & -     & \cite{NGTS1}              \\
NGTS-10    & 14.51 & 13.61 & 0.7  & 0.7  & 4600& 13.51 & 0.7669   & 2    & -     & \cite{NGTS10}             \\
NGTS-15    & 14.67 & 13.96 & 0.9  & 1.0  & 5600& 12.33 & 3.2762   & 1    & -     & \cite{NGTS15_18}         \\
NGTS-17    & 14.41 & 13.77 & 1.3  & 1.0  & 5650& 13.9  & 3.2425   & 2    & -     & \cite{NGTS15_18}           \\
NGTS-23    & 14.13 & 13.59 & 1.2  & 1.0  & 6057& 14.2  & 4.0764   & 4    & -     & \cite{NGTS23_25}         \\
NGTS-29 & 10.51 & 10.01 & 1.0 & 1.0 & 5730 & 9.61 & 69.3368 & 3 & 6 & \cite{ngts29}\\
NGTS-3 A   & 14.67 & 13.98 & 0.9  & 1.0  & 5600& 16.59 & 1.6754   & 4    & -     & \cite{NGTS3}              \\
NGTS-4     & 13.14 & 12.37 & 0.8  & 0.8  & 5143& 3.18  & 1.3374   & 4    & -     & \cite{NGTS4}              \\
NGTS-6     & 14.24 & 13.17 & 0.8  & 0.8  & 4730& 14.86 & 0.8821   & 2    & -     & \cite{NGTS6}              \\
TOI-1011 & 8.94 & 8.24 & 0.9 & 0.9 & 5475 & 1.45 & 2.4705 & 4 & - & \cite{toi1011} \\
TOI-1221   & 10.49 & 10.06 & 1.0  & 0.9  & 5592& 2.91  & 91.6828  & 42   & 12    & \cite{toi1221}            \\
TOI-163    & 11.47 & 10.87 & 1.6  & 1.4  & 6495& 16.69 & 4.2313   & 39   & 6     & \cite{toi163}            \\
TOI-1937 A & 13.18 & 12.49 & 1.1  & 1.1  & 5814& 13.98 & 0.9467   & 9    & -     & \cite{toi2803}            \\
TOI-201    & 9.07  & 8.58  & 1.3  & 1.3  & 6394& 11.3  & 52.9782  & 36   & 24    & \cite{TOI201}             \\
TOI-206    & 14.94 & 12.43 & 0.3  & 0.3  & 3383& 1.3   & 0.7363   & 42   & 6     & \cite{TOI500_1}             \\
TOI-2184   & 12.25 & 11.41 & 2.9  & 1.5  & 5966& 11.4  & 6.9068   & 39   & 12    & \cite{toi2184}            \\
TOI-220    & 10.47 & 9.69  & 0.9  & 0.8  & 5298& 3.03  & 10.6953  & 37   & 12    & \cite{toi220}             \\
TOI-2338   & 12.48 & 11.7  & 1.1  & 1.0  & 5581& 11.21 & 22.654   & 3    & 12    & \cite{toi2338}            \\
TOI-2368   & 12.49 & 11.7  & 0.8  & 0.9  & 5360& 10.84 & 5.175    & 8    & -     & \cite{TOI2368}            \\
TOI-2416   & 13.02 & 12.43 & 1.2  & 1.1  & 5808& 9.86  & 8.2755   & 15   & -     & \cite{TOI2416}            \\
TOI-2459   & 10.77 & 9.39  & 0.7  & 0.7  & 4195& 2.95  & 19.1047  & 5    & 18    & \cite{toi2459}            \\
TOI-2529   & 11.53 & 10.67 & 1.7  & 1.1  & 5802& 11.54 & 64.5949  & 11    & -     & \cite{TOI2529}            \\
TOI-2589   & 11.41 & 10.72 & 1.1  & 0.9  & 5579& 12.11 & 61.6277  & 6    & 12    & \cite{toi2338}             \\
TOI-269    & 14.37 & 12.3  & 0.4  & 0.4  & 3514& 2.77  & 3.6977   & 19   & 12    & \cite{toi269}             \\
TOI-2803 A & 12.54 & 12.07 & 1.2  & 1.1  & 6280& 18.11 & 1.9623   & 3    & 6     & \cite{toi2803}            \\
TOI-2818   & 11.94 & 11.39 & 1.2  & 1.0  & 5721& 15.28 & 4.0397   & 5    & 12    & \cite{toi2803}            \\ 
TOI-470    & 11.17 & 10.7  & 0.8  & 0.9  & 5190& 4.34  & 12.1915  & 3    & 6     & \cite{toi470}            \\
TOI-481    & 9.97  & 9.39  & 1.7  & 1.1  & 5735& 11.1  & 10.3311  & 26   & 12    & \cite{toi481}             \\
TOI-540    & 14.82 & 11.51 & 0.2  & 0.2  & 3216& 0.9   & 1.2391   & 5    & 12    & \cite{TOI540}             \\
TOI-622    & 8.99  & 8.56  & 1.4  & 1.3  & 6400& 9.24  & 6.4025   & 8    & 12    & \cite{toi622}            \\
TOI-640    & 10.51 & 10.04 & 2.1  & 1.5  & 6460& 19.85 & 5.0038   & 5    & 18    & \cite{toi640}             \\
TOI-813    & 10.36 & 9.85  & 1.9  & 1.3  & 5907& 6.71  & 83.8911  & 40   & 6     & \cite{TOI813}             \\
TOI-858 B  & 11.18 & 10.64 & 1.3  & 1.1  & 5842& 14.07 & 3.2797   & 6    & 6     & \cite{toi858}             \\
TOI-871    & 10.57 & 9.76  & 0.7  & 0.8  & 4929& 1.66  & 14.3626  & 5    & 12    & \cite{TOI871}             \\
WASP-100   & 10.8  & 10.32 & 1.6  & 0.8  & 6900& 14.91 & 2.8494   & 41   & 6     & \cite{wasp100}            \\
WASP-101   & 10.34 & 9.79  & 1.3  & 1.4  & 6380& 16.03 & 3.5857   & 2    & 6     & \cite{wasp100}            \\
WASP-119   & 12.31 & 11.6  & 1.2  & 1.0  & 5650& 15.69 & 2.4998   & 27   & 6     & \cite{WASP126b}           \\
WASP-120   & 10.96 & 10.58 & 1.9  & 1.4  & 6450& 16.51 & 3.6113   & 4    & 12    & \cite{wasp120}            \\
WASP-121   & 10.51 & 10.06 & 1.5  & 1.4  & 6776& 19.65 & 1.2749   & 6    & 12    & \cite{wasp121}            \\
WASP-126   & 10.99 & 10.61 & 1.2  & 1.1  & 5633& 10.8  & 3.2888   & 33   & 6     & \cite{WASP126b}           \\
WASP-159   & 12.84 & 11.88 & 2.1  & 1.4  & 6120& 15.47 & 3.8404   & 4    & 6     & \cite{wasp159}            \\
WASP-160 B & 13.04 & 12.34 & 0.9  & 0.9  & 5298& 12.22 & 3.7685   & 4    & -     & \cite{WASP160B}           \\
WASP-168   & 12.12 & 11.32 & 1.1  & 1.1  & 6000& 16.81 & 4.1537   & 7    & 24    & \cite{wasp159}            \\
WASP-23    & 12.54 & 11.76 & 0.8  & 0.8  & 5150& 10.78 & 2.9444   & 5    & 24    & \cite{WASP23}             \\
WASP-61    & 12.49 & 11.72 & 1.6  & 1.8  & 6250& 15.8  & 3.8559   & 2    & 6     & \cite{wasp61}             \\
WASP-62    & 10.21 & 9.71  & 1.2  & 1.1  & 6230& 14.8  & 4.412    & 36   & 12    & \cite{wasp61}             \\
WASP-63    & 11.15 & 10.44 & 1.9  & 1.3  & 5550& 15.8  & 4.3781   & 5    & 24    & \cite{wasp61}             \\
WASP-64    & 12.7  & 12.05 & 1.1  & 1.0  & 5400& 14.25 & 1.5733   & 5    & 12    & \cite{wasp64}            \\
WASP-79    & 10.04 & 9.68  & 1.5  & 1.4  & 6600& 17.15 & 3.6624   & 4    & 6     & \cite{wasp79} \\    
\hline
\end{tabular}
\caption{Confirmed transiting planet host stars with only one confirmed planet to date in the \plato\, LOPS2 Field. Stellar parameters and the planet radius and the orbital period of each confirmed planet are obtained from the Exoplanet Archive \citep[][accessed on 9 October 2024]{PSCompPars} and the number of \plato\, cameras the stars will be observed with are obtained from the LOPS2 PIC. For planets that are currently not in the LOPS2 PIC the number of cameras is marked with a minus.}
\label{tab:transit_planets}
\end{table*}

\begin{table*}
    \centering
    \begin{tabular}{llllllllll}
    \hline
         Host Star & VMag & R$_S$ & M$_S$ & Teff& \# Planets & Max Period & TESS & PLATO & Paper\\
         &  &  (R$_\odot$) & (M$_\odot$) & (K) & & (days)  & Sectors & Cameras \\
        \hline
          HD 23472 & 9.73 & 0.7 & 0.7 & 4684 & 5 & 29.80 & 14 & 6 & \cite{HD23472_2, HD23472} \\ 
          HD 28109 & 9.42 & 1.4 & 1.3 & 6120 & 3 & 84.26 & 42 & 6 & \cite{2022MNRAS.515.1328D} \\ 
          LHS 1678 & 12.6 & 0.3 & 0.3 & 3490 & 3 & 4.97 & 4 & 6 & \cite{LHS1678_1}, \cite{LHS1678_2} \\
          TOI-1338 A & 11.72 & 1.3 & 1.1 & 6050 & 2 & 215.5$^1$ & 41 & 18 &\cite{toi1338_1, toi1338_2}\\ 
          TOI-199 & 10.70 & 0.8 & 0.9 & 5255 & 2 & 273.69$^2$ & 40 & 18 &\cite{TOI199} \\ 
          TOI-216 & 12.32 & 0.7 & 0.8 & 5026 & 2 & 34.53 & 42 & 6 & \cite{TOI216}\\
          TOI-2525 & 14.22 & 0.8 & 0.8 & 5096 & 2 & 49.25 & 38 & - & \cite{TOI2525}\\
          TOI-270 & 12.60 & 0.4 & 0.4 & 3506 & 3 & 11.40 & 5 & 12  & \cite{TOI270}\\
          TOI-286 & 9.87 & 0.8 & 0.8 & 5152 & 2 & 39.36 & 42 & 12 & \cite{TOI286} \\
          TOI-431 & 9.12 & 0.7 & 0.8 & 4850 & 3 & 12.46 & 3 & 6 & \cite{TOI431}\\
          TOI-451 & 10.94 & 0.9 & 1.0 & 5550 & 3 & 16.36 & 3 & 6 & \cite{TOI451}\\
          TOI-4562   & 12.14 & 1.1  & 1.2  & 6096& 2 & 3990$^3$ & 42   & 12    & \cite{toi4562, TOI4562_2} \\
          TOI-500 & 10.54 & 0.7 & 0.7 & 4440 & 4 & 61.3 & 9 & 24 & \cite{TOI500_1}, \cite{TOI500_2}\\
          TOI-700 & 13.15 & 0.4 & 0.4 & 3459 & 4 & 37.42 & 36 & 12 &  \cite{TOI700} \\
          TOI-712 & 10.84 & 0.7 & 0.7 & 4622 & 3 & 84.84 & 30 & 12 & \cite{toi712}\\
         \hline
    \end{tabular}
    \caption{Confirmed multiplanetary systems with at least one transiting planet in the \plato\, LOPS2 Field. Stellar parameters and the planet radius and the orbital period of each confirmed planet are obtained from the Exoplanet Archive \citep[][accessed on 9 October 2024]{PSCompPars} and the number of \plato\, cameras the stars will be observed with are obtained from the LOPS2 PIC. For planets that are currently not in the LOPS2 PIC the number of cameras is marked with a minus.\\
    $^1$ TOI-1338 A\,c with a period of 215.5 days was only detected through radial velocity but not transit.\\
    $^2$ TOI-199\,c with a period 273.69 days was detected through TTV but not transit.\\
    $^3$ TOI-4562\,c with a period 3390 days was detected through TTV but not transit.}
    \label{tab:multis}
\end{table*}

\begin{table*}
\begin{tabular}{lllllllllll}
\hline
System & Vmag  & R$_S$ & M$_S$ & Teff  & Period & Radius & Mass & TESS & PLATO & Paper \\
    &   & (R$_\odot$) & (M$_\odot$) & (K)  & (days) & (R$_J$) & (M$_J$) & Sectors & Cameras \\
\hline
HIP 33609 & 7.28 & 1.9 & 2.4 & 10400 & 39.471814   & 1.58 & 68.0 & 8 & - & \cite{HIP33609} \\
TOI-569  & 10.17  & 1.5 & 1.1 & 5705 & 6.55604     & 0.75 & 64.1 & 5 & 12 &\cite{TOI569_1406} \\
TOI-811    & 11.41  & 1.2 & 1.1 & 6107 & 25.16551    & 1.262 & 55.3 & 5 & 12 &\cite{TOI811} \\
TOI-1406   & 12.07  & 1.3 & 1.2 & 6290 & 10.57398    & 0.86  & 46.0 & 7 & 24 & \cite{TOI569_1406} \\
TOI-2490 & 12.18 & 1.1 & 1.0 & 5558 & 60.33 & 1.00 & 73.6 & 3 & 6 & \cite{TOI2490}\\
HATS-70    & 12.57  & 1.9 & 1.8 & 7930 & 1.8882378   & 1.384 & 12.9 &6 & - & \cite{hats70}\\
KELT-25    & 9.84   & 2.3 & 2.2 & 8100 & 4.401131    & 1.642 & $\sim$64 & 6 & - & \cite{KELT25}\\
\hline
\end{tabular}
\caption{Transiting Brown Dwarfs in the LOPS2 Field. }
\label{tab:brown_dwarfs}
\end{table*}

\begin{table*}
\begin{tabular}{llllllllll}
\hline
Host Star   & VMag & R$_S$ & M$_S$ & Teff & M sin(i)  & Period  & TESS & PLATO & Paper \\
&  &  (R$_\odot$) & (M$_\odot$) & (K) & (M$_\oplus$)  & (days)  & Sectors & Cameras \\
\hline
DMPP-3 A  & 9.07     & 0.9     & 0.9      & 5138   & 2.58       & 6.6732     & 41      &  6 &\cite{DMPP_3A}             \\
GJ 2056   & 10.37    & 0.7     & 0.6      & 4070   & 16.2       & 69.971     & 6    & 6   & \cite{GJ_2056}        \\
GJ 3341   & 12.06    & 0.4     & 0.5      & 3526   & 6.6        & 14.207     & 3     & 6           
 & \cite{GJ3341}\\
HD 23127  & 8.58     & 1.5     & 1.2      & 5843   & 485.33     & 1211.17    & 9    & 6  & \cite{HD23127}         \\
HD 25171  & 7.77     & 1.2     & 1.1      & 6125   & 290.81     & 1802.29    & 36    & 6    & \cite{HD25171}       \\
HD 27442  & 4.44     & 3.2     & 1.2      & 4846   & 495.8      & 428.1      & 26    &  - & \cite{HD27442}             \\
HD 27631  & 8.26     & 0.9     & 0.9      & 5737   & 495.81     & 2198.14    & 4     &   - & \cite{HD27631}            \\
HD 28254  & 7.69     & 1.5     & 1.1      & 5664   & 1207.75    & 1333.0     & 6     & 12 & \cite{HD43197_1_HD28254}          \\
HD 29399  & 5.79     & 4.5     & 1.2      & 4845   & 498.99     & 892.7      & 40    &    - & \cite{HD29399}           \\
HD 29985  & 9.98     & 0.6     & 0.8      & 4678   & 1739.47    & 20608.1405 & 3    & 6 & \cite{HD29985_HD56957_HIP19976}           \\
HD 30669  & 9.12     & 0.9     & 0.9      & 5400   & 149.37     & 1684.0     & 2     & 6 & \cite{HD30669}           \\
HD 33283  & 8.05     & 2.0     & 1.4      & 5935   & 104.57     & 18.1991    & 2     & 6 & \cite{HD33283}           \\
HD 38283  & 6.69     & 1.5     & 1.4      & 5981   & 127.13     & 363.2      & 40    & 6 & \cite{HD38283}           \\
HD 47536  & 5.25     & 23.5    & 2.1      & 4380   & 2330.0     & 712.13     & 4     &   - & \cite{HD47536}            \\
HD 48265  & 8.05     & 1.9     & 1.3      & 5733   & 484.69     & 778.51     & 9     & 24 & \cite{HD48265}          \\
HD 55696  & 7.95     & 1.5     & 1.3      & 6012   & 1357.13    & 1827.0     & 6     & 12 & \cite{HD55696}          \\
HD 56957  & 7.57     & 1.8     & 1.0      & 5743   & 1880.91    & 5859.0899  & 7     & 12 & \cite{HD29985_HD56957_HIP19976}          \\
HD 63765  & 8.1      & 0.8     & 0.6      & 5449   & 168.45     & 358.0      & 12    & 18 & \cite{HD63765}         \\
HD 64121  & 7.43     & 5.4     & 1.6      & 5078   & 813.64     & 623.0      & 12    &   - & \cite{HD64121_HD69123}            \\
HD 69123  & 5.77     & 7.7     & 1.7      & 4842   & 982.09     & 1193.3     & 7     &  -    & \cite{HD64121_HD69123}          \\
HD 70642  & 7.17     & 1.0     & 1.1      & 5732   & 1239.53    & 2124.54    & 7     & 12 & \cite{HD70642}          \\
HIP 19976 & 10.48    & 0.8     & 0.7      & 4316   & 21.61      & 50.659     & 4     & 6 & \cite{HD29985_HD56957_HIP19976}          \\
HIP 35965 & 10.18    & 0.7     & 0.8      & 4836   & 2589.98    & 26125.565  & 11    & 6  & \cite{HD29985_HD56957_HIP19976}         \\
Kapteyn's Star   & 8.86     & 0.3     & 0.3      & 3550   & 7.0        & 121.54     & 6     & 12  &  \cite{Kapteyn_planets}     \\
\hline  
\end{tabular}
\caption{Confirmed radial velocity planet host stars with only one confirmed planet to date in the \plato\, LOPS2 Field. Stellar parameters and the planet radius and the orbital period of each confirmed planet are obtained from the Exoplanet Archive \citep[][accessed on 24. May 2024]{PSCompPars} and the number of \plato\, cameras the stars will be observed with are obtained from the LOPS2 PIC. For planets that are currently not in the LOPS2 PIC the number of cameras is marked with a minus.}
\label{tab:rv_planets}
\end{table*}

\begin{table*}
\begin{tabular}{lllllllll}
\hline
Host Star   & \# Planets & VMag & R$_S$ & M$_S$ & Teff & TESS & PLATO &   Paper \\
  &  &  & (R$_\odot$) & (M$_\odot$) & (K) & Sectors & Cameras &  \\
\hline
GJ 163     & 3        & 11.79    & 0.4     & 0.4      & 3500 & 5 & 12 & \cite{GJ163}  \\
HD 25912   & 2        & 8.2      & 1.0     & 1.1      & 5921  & 5 & 12 &  \cite{HD29985_HD56957_HIP19976}\\
HD 27894   & 3        & 9.36     & 0.9     & 0.8      & 4875 & 29 & 6 & \cite{HD27894_1, HD27894_2}  \\
HD 30177   & 2        & 8.41     & 1.0     & 1.0      & 5580  & 34 & 12 & \cite{HD30177_1, HD30177_2} \\
HD 39194   & 3        & 8.09     & 0.8     & 0.7      & 5205 & 42 & 6 & \cite{HD39194} \\
HD 40307   & 5        & 7.17     & 0.7     & 0.8      & 4956 & 40 & 6 & \cite{HD40307_1, HD40307_2}  \\
HD 41004 A & 2        & 8.65     & 0.8     & 1.0      & 5310  & 7 & 24 & \cite{HD41004A}\\
HD 41004 B & 2        & 8.61     & 1.0     & 0.4      & 5036  & 7 & - & \cite{HD41004B} \\
HD 43197   & 2        & 8.98     & 1.0     & 1.0      & 5508  & 3 & 12 & \cite{HD43197_1_HD28254, HD29985_HD56957_HIP19976} \\
HD 45184   & 2        & 6.38     & 1.1     & 1.0      & 5869  & 3 & 12 & \cite{HD45184}\\
HD 45364   & 2        & 8.08     & 0.9     & 0.8      & 5466  & 3 & 12 & \cite{HD45364} \\
HD 47186   & 2        & 7.63     & 1.1     & 1.0      & 5657  & 4 & 6 & \cite{HD47186} \\
HD 50499   & 2        & 7.21     & 1.4     & 1.2      & 6102  & 6 & 12 & \cite{HD50499_1, HD50499_2} \\
HD 51608   & 2        & 8.17     & 0.9     & 0.8      & 5358  & 29 & 24 & \cite{HD45184} \\
HD 65216   & 2        & 7.97     & 0.9     & 0.9      & 5612  & 37 & 12 &  \cite{HD65216_1, HD65216_2}\\
TOI-1338 A & 2        & 11.72    & 1.3     & 1.1      & 6050  & 41 & 18 & \cite{toi1338_1, toi1338_2}\\
TOI-431    & 3        & 9.12     & 0.7     & 0.8      & 4850  & 3 & 6 &  \cite{TOI431}\\
TOI-500    & 4        & 10.54    & 0.7     & 0.7      & 4440  & 9 & 24 & \cite{TOI500_1, TOI500_2}\\
bet Pic    & 2        & 3.85     & 2.0     & 1.8      & 7890  & 9 & - & \cite{betapic_1, betapic_2}\\
\hline
\end{tabular}
\caption{Confirmed multiplanetary systems in the LOPS2 field with at least one planet that has only been detected through radial velocity. Stellar parameters and the planet radius and the orbital period of each confirmed planet are obtained from the Exoplanet Archive \citep[][accessed on 24. May 2024]{PSCompPars} and the number of \plato\, cameras the stars will be observed with are obtained from the LOPS2 PIC. For planets that are currently not in the LOPS2 PIC the number of cameras is marked with a minus.}
\label{tab:rv_multis}
\end{table*}

%%%%%%%%%%%%%%%%%%%%%%%%%%%%%%%%%%%%%%%%%%%%%%%%%%

% Don't change these lines
\bsp	% typesetting comment
\label{lastpage}
\end{document}